\documentclass[aps,prl,preprint,groupedaddress]{revtex4}
\usepackage{epsfig}
\usepackage{latexsym}
\usepackage{graphics}
\usepackage{graphicx}
\begin{document}
%
\newcommand{\ds}{\displaystyle}
\title{A six-quark dressed-bag description
 of $np \to d \gamma $ radiative capture}
\author{M. M. Kaskulov}
 \email{kaskulov@pit.physik.uni-tuebingen.de}
\affiliation{Physikalisches Institut, Universit\"at  T\"ubingen,
		 D-72076 T\"ubingen, Germany}
\affiliation{ Institute of Nuclear Physics, Moscow State University,
  119899 Moscow, Russia}
\author{V. I. Kukulin}
\affiliation{Institut f\"ur Theoretische Physik, Universit\"at
  T\"ubingen, D-72076 T\"ubingen, Germany}
\affiliation{ Institute of Nuclear Physics, Moscow State University,
  119899 Moscow, Russia}
\author{P. Grabmayr}
\affiliation{Physikalisches Institut, Universit\"at  T\"ubingen,
		 D-72076 T\"ubingen, Germany}

\begin{abstract}
The radiative capture process $n p \to d \gamma$ is considered
 within the framework of a recently developed six-quark dressed-bag model for
 the nucleon-nucleon interaction. The calculations presented here include both
 the nucleon current and the meson-exchange current contributions. The latter
 uses short-range hadronic form factors for the pion exchange currents
 consistent with the soft cut-off parameter $\Lambda_{\pi NN}$ from the
 $NN$-potential.  Contributions of the pion exchange current and
 $\Delta$-isobar current to the total cross section still cannot explain the
 discrepancy between the theoretical and experimental cross sections.
 Possibilities for new types of meson exchange currents associated with
 chiral fields inside multi-quark dressed-bag states in nuclei are discussed.
%
%
\end{abstract}
\maketitle

\section{Introduction and motivation}
\label{intro}
Radiative capture of thermal neutrons by hydrogen, $^1$H$(n,\gamma)^2$H, is
 now a well-studied reaction which is of prime importance in both
 astrophysical applications and in fundamental nuclear physics.  This process
 is one of the simplest classical nuclear interactions involving an
 electromagnetic (e.-m.) ``probe'', and thus it is doubly interesting due to
 its close relationship to $180^{\circ}$ inelastic electron-deuteron
 scattering near threshold, as well as to deuteron photodisintegration $\gamma
 d \to n p$.  The primary interaction is of e.-m. nature, and thus it is well
 understood; the leading effects of strong interaction occurring between two
 nucleons are also well characterised. Therefore, observables for this
 reaction, if measured with sufficient precision, can provide a sensitive test
 to phenomena associated with, for example, the role of sub-nucleonic degrees
 of freedom.

The numerous measurements done in previous years have demonstrated that the
 accepted values for $\sigma_{np}$($^1$H) are free of systematic errors and
 without masking due to competing processes. The modern accepted value of the
 $np \to d \gamma$ cross section for thermal neutron capture~\cite{CWC} at a
 neutron velocity of 2200~m/sec is
\begin{equation}
\label{Exp_np_val}
\sigma_{np}(^1 \mbox{H}) = 334.2 \pm 0.5 
\,\, \mbox{mb}.
\end{equation}
We will compare this experimental result  with theory. 

For many years a 10{\%} discrepancy has existed between experiment and
 theoretical calculations for the total radiative thermal $np$ capture cross
 section, $n p \to d \gamma$.  The discrepancy was first pointed out by
 Austern~\cite{Austern1953} as early as in 1953 and has been repeatedly
 verified in the literature since then. The successful explanation of the
 cross section in terms of meson exchange currents (MEC) given three decades
 ago by Riska and Brown~\cite{Riska_Brown} was considered as one of the
 cornerstones of mesonic degrees of freedoms in nuclear physics. Using a
 realistic hard-core wave function for the deuteron, these authors computed
 the two diagrams with one-pion exchange (OPE) initially suggested in 1947 by
 Villars~\cite{Villars} plus $\omega$ and $\Delta$ resonance
 diagrams. Although suspected since the Yukawa force was introduced, the work
 of Riska and Brown was the first evidence for the explicit role of mesons,
 in particular that of pions, in nuclear interactions. In recent years the
 status of meson-exchange currents in the deuteron, including the $np \to d
 \gamma$ reaction, has been discussed exhaustively by
 Arenh\"ovel~\cite{Arenh95}, Mathiot~\cite{Mathiot89} and
 Riska~\cite{Riska89}.

However, apart from quantitative calculations showing the sensitivity of the
 cross section to the $NN$ wave function (at a level of
 1.5$\%$)~\cite{Gari73}, short-range contributions to exchange currents ($\pi
 NN$ form factor, $\rho$-exchange currents) have been studied only
 recently. These contributions appear to be of some importance, especially in
 the light of current pictures of the quark substructure of the nucleon.

In the framework of chiral perturbation theory ($\chi$PT) it has been shown
 that the short-range interaction is suppressed in the exchange current
 contributions~\cite{Rho1991}.  From this the conclusion was
 reached~\cite{Kubodera1978}, that only {\it soft} one-pion exchange terms
 contribute to the two-body current. This means that all short-range hadronic
 processes and higher order corrections are cancelled out to rather high
 accuracy for the magnetic transitions leaving only the pion degree of freedom
 --- ``{\it a chiral filter}''.  This important conclusion shifted the problem
 of sensitivity of all the MEC-contributions to the values of cut-off
 parameters $\Lambda_{\pi NN}, \Lambda_{\rho NN}, \Lambda_{\pi N\Delta}$
 etc. used in current theories for MEC. In fact, all modern one-boson exchange
 (OBE)-models suggested up to date for the description of the $NN$-interaction
 employ rather high cut-off values $\Lambda_{\pi NN} \simeq 1.5$~GeV~\cite{1a}
 and thus the currently accepted approaches for MEC also include similarly
 high values of cut-off parameters. Meanwhile, all consistent theories for
 $\pi-N$ interactions and also numerous experimental data result unambiguously
 in {\it soft} cut-off parameters around 600~MeV/c~\cite{2a}.  However, these
 soft cut-off values cannot be matched with the quantitative description of
 the $NN$-interaction within the OBE-model~\cite{3a,3aa}. Moreover these soft
 cut-offs, being in agreement with the above chiral filter ideas, lead
 unavoidably to a significant reduction of all the MEC-contributions to the $n
 p \to d \gamma$ process, and thus there appears again some visible
 disagreement between the experimental data and current theoretical
 estimations for MEC-contributions.

In the present work we try to solve this puzzle by use of a new $NN$
 microscopic force model developed recently jointly by the Moscow-T\"ubingen
 (MT) group~\cite{2aa,4a}. Contrary to meson-exchange models, the new model
 includes soft cut-off parameter values only, and thus it is in agreement with
 the above ``chiral filter'' ideas while providing a quantitative and accurate
 description of $NN$-data~\cite{4a}.

Another specific feature of the new force model is the fact that mesonic and
 quark degrees of freedom are included into the model in a consistent and
 explicit form, so that the total wave function of the $NN$ system is described
 as a Fock row:
\begin{equation}
\label{Fock2}
\Psi(NN) = \left(
\begin{array}{c}
\Psi_{NN} \\
\Psi_{6q + \sigma} \\
\Psi_{6q + 2 \pi} \\
\Psi_{6q + \rho} \\
\vdots
\end{array}
\right)
\end{equation}
here the second and higher components are of quark-mesonic nature and are
 essential at short and intermediate ($r < 1$~fm) ranges. On the other hand,
 it would be very interesting to check whether there is a non-negligible
 contribution to the $np \to d \gamma$ reaction at low energies from the
 region $r < 1$~fm, especially for meson exchange-current contributions.
 Another interesting problem to study here is the role of ``inner'' tensor
 force introduced within the new $NN$-model~\cite{2aa,4a}.  This specific
 short-range tensor force is based on the symmetry of the six-quark wave
 functions and leads also to some enhancement of the $D$-wave function of the
 deuteron in the asymptotic region. This modification should affect the
 radiative capture cross section.

To answer the above key questions, we present in this paper a consistent
 {\it ab initio} calculation for the cross section of the $np \to d \gamma$
 reaction with thermal neutrons using the new $NN$-force model.

The structure of the article is the following: section~2 is dedicated to the
 general features of $np$-radiative capture and its relationship to the
 underlying $NN$-force model. In section~3 we present a brief description of
 the new force model. The deuteron structure emerging from the new $NN$-model
 is discussed in section~4. The central section~5 is devoted to the formalism
 for e.-m. currents within the quark-meson force model while the calculational
 results are presented in section~6. In the concluding section~7 we 
 discuss the results obtained and outline general perspectives of
 our new approach.

\section{General features of $np$ capture}
\label{sec:2}

The main features of low energy $np$ capture are generally supposed to be well
 understood~\cite{Blatt}. We list here the basic points specific to this
 process.

At threshold the neutron is captured predominantly from an $l=0$ state.  This
 is to be expected due to dominant $S$-wave component of the deuteron wave
 function. This statement was verified by explicit calculations reported by
 Adler~\cite{Adler1} and Jankus~\cite{Jankus1}. Moreover, it was found that
 the dominant transition proceeds from the $np$ $^{1}S_0$ initial state, while
 the $^{3}S_1$ state is assumed to be of little importance. Since the deuteron
 ground state is dominantly $^{3}S_1$, the transition involves a change in
 total angular momentum of $\Delta J =1$ and thus it is a $M1$ transition.

The transition from the $^{1}S_0$ to the $^{3}S_1$ component of the deuteron
 wave function dominates the impulse approximation~\cite{Bethe}.  Since the
 continuum $^{1}S_0$ state is of isovector while the deuteron is of isoscalar
 nature, the dominant $^{1}S_0 \to d$ transition has an isovector character
 and thus the transition involves an isospin-flip amplitude; the emitted
 photon has $I=1$. The isovector magnetic moment for the two-nucleon system,
 $(\mu_p - \mu_n)$, can be used to simplify the calculations. Similarly the
 $^{3}S_1 \to d$ transition is isoscalar and involves the isoscalar magnetic
 moment, $(\mu_p + \mu_n)$, and isoscalar charge.

The problem under question is closely related to both deuteron photo- and
 electro-disintegration. Indeed, the magnetic photo-disintegration is merely
 the time-reversed process to the $np \to d \gamma$ reaction and thus involves
 precisely the same matrix elements and transition currents as the latter
 reaction. The deuteron electro-disintegration requires a broader knowledge of
 the transition current process; more precisely, the relevant magnetic
 transition current involves a value of momentum-transfer squared,
 $q^{\mu}q_{\mu} = q^2$. The radiative capture process is accompanied by
 emission of a $\gamma$-quant with $q^2= \omega_{\gamma}^2 - {\bf q}^2 = 0$
 where $|{\bf q}|$ equals the deuteron binding energy, $|{\bf q}| =
 \epsilon_d$.  Thus, the current we are dealing with here is merely a special
 case of the currents involved in electro-disintegration.

The expression for the cross section can be divided into the phase space and
 kinematical factors and into the current transition matrix element which
 describes the dynamics:
\begin{equation}
d \, \sigma_{fi} = \frac{\delta (P_{f} - P_{i})}
			{\left| {\bf v}_n -{\bf v}_p \right|}
 \frac{1}{(2 \pi)^2} \frac{1}{2 \omega_{\gamma}}
\frac{M_{p}M_{n}}{E_p E_n} \frac{M_d}{E_d}
\left| \mathcal{M}_{fi} \right|^2 d^3 {\bf p}_d d^3 {\bf q}_{\gamma} 
\end{equation}
where the transition amplitude takes the form:
\begin{equation}
\mathcal{M}_{fi} =  \langle{\psi_d} | \vec{j}({\bf q})
 \cdot \vec{\epsilon}_{\lambda} | \psi_{pn} \rangle
\end{equation}
with $\vec{j}({\bf q})$ for the hadronic e.-m. current and
 $\vec{\epsilon}_{\lambda}$ stands for the photon polarisation
 vector. $\psi_d$ and $\psi_{pn}$ are the initial and final state wave
 functions.

\section{The effective $NN$-interaction in the dressed-bag state model}
\label{sec:effint}

The long-range part of the $NN$-potential is certainly the most accurately
 known and it is generated from the exchange of $\pi$-mesons.  This part of the
 potential provides a rather good description of the static properties of the
 deuteron, and gives rise to a noticeable $D$-wave component in the deuteron
 wave function. The structure of the $NN$-interaction at intermediate and
 especially at short distances is still not well understood.  In conventional
 models it is assumed to be generated from two-and three-pion exchange,
 although such a construction is very complicated.  Moreover, this part is
 fundamental in the treatment of the nucleon structure since the $\pi$-mesons
 are regarded as the Goldstone bosons associated with spontaneous breaking of
 chiral symmetry. A complete account of the physics related to the pion
 degrees of freedom can be found in ref.~\cite{Ericson_Weise}.

The most natural way to take into account the multiple $\pi$-exchange is to
 use dispersion relation techniques.  The important ingredients are the
 correlated and uncorrelated $2\pi$-exchange contributions. The part which
 comes from uncorrelated $2\pi$-exchange involves isobar components in the
 intermediate states; both crossed and uncrossed diagrams contribute. Another
 part comes from pion rescattering and is included through the $\pi \pi$
 scattering amplitude to the $t$-channel.  It was assumed in previous
 years~\cite{3a}, that the $2\pi$-exchange amplitude with $\pi$$-$$\pi$
 resonance-like interaction in the $s$-channel and $\Delta$-isobar
 intermediate states can be parametrised, in first order, by the exchange of
 heavier scalar-isoscalar mesons.  This gave rise to the concept of the
 scalar-isoscalar $\sigma$~meson exchange.  The $NN$-potential constructed in
 this way is momentum- (and/or energy-) dependent. This is certainly not very
 surprising since at very short distances the potential has no reason to be
 local. In the literature we can find several modern parametrisations of the
 $NN$-potential (see e.g. refs.~\cite{1a,3a,3aa,5a}); they may be classified
 according to their description of the medium and short-range part. However,
 it has been demonstrated independently by a few groups that the exchange of a
 correlated $2\pi$ pair in the scalar-isoscalar channel cannot give any
 significant attraction between two nucleons sufficient to couple them to a
 deuteron or to produce nuclear binding~(see e.g. ref.~\cite{6a}). This
 failure has perhaps a fundamental character for our understanding of the
 nuclear force, and thus could mean some deep revision of the force at
 intermediate distances.

 Another fundamental problem with the current status of the force is the
 short-range behaviour of the $NN$-interaction in terms of meson-nucleon form
 factors. In fact, at short distances $(r < 0.8$ fm), all these potentials are
 regularised in either a completely phenomenological way (e.g. as in the case of
 the Paris $NN$-potential) or by phenomenological form factors at each
 meson-nucleon vertex with cut-off parameters fitted to $NN$-phase
 shifts~\cite{3a}.  This short-range part of the $NN$-potential is indeed very
 crucial in nuclear physics since it gives rise to a strong repulsion between
 the nucleons and leads in particular to the saturation of nuclear matter in
 the usual non-relativistic many-body formalism. However the cut-off parameter
 values $\Lambda_{\pi NN}$, $\Lambda_{\rho NN}$, $\Lambda_{\pi N \Delta}$
 etc., accepted in current OBE-models for $NN$-potentials, e.g. $\Lambda_{\pi
 NN} \simeq 1.5-1.7$~GeV, exceed the respective cut-off parameter values
 derived independently from {\it all microscopic} and dynamical models for
 $\pi N$ interaction, as well as the values inferred from experiment,
 $\Lambda^{\mbox{\tiny mono}}_{\pi NN} \simeq 0.5 - 0.7$~GeV.

 There is another serious problem with the short-range $NN$-interaction which
 is attributed mainly to $\omega$-meson exchange. The $\omega NN$ coupling
 constant $g^{2}_{\omega NN}/4 \pi$, providing a good fit to $NN$-scattering
 data, i.e. $g^{2}_{\omega NN}/4 \pi = 13 \div 15$, exceeds three times the
 value dictated by $SU(3)$-symmetry~\cite{2a}, while all remaining couplings
 are still in agreement with this symmetry. Moreover, Feshbach has
 demonstrated~ that the $SU(3)$ symmetry should be a rather good one in this
 region (see the detailed discussion of the above mentioned disagreements and
 inconsistencies in OBE-models in ref.~\cite{2a}).  These strong disagreements
 peculiar to the current OBE-models require a search for another model for the
 short-range behaviour of $NN$-potentials. Such an alternative model for the
 $NN$-force has been developed recently~\cite{2aa,4a} by the Moscow-T\"ubingen
 group.

\begin{figure}[t]\centering
\includegraphics[scale=0.5, angle=-90.]{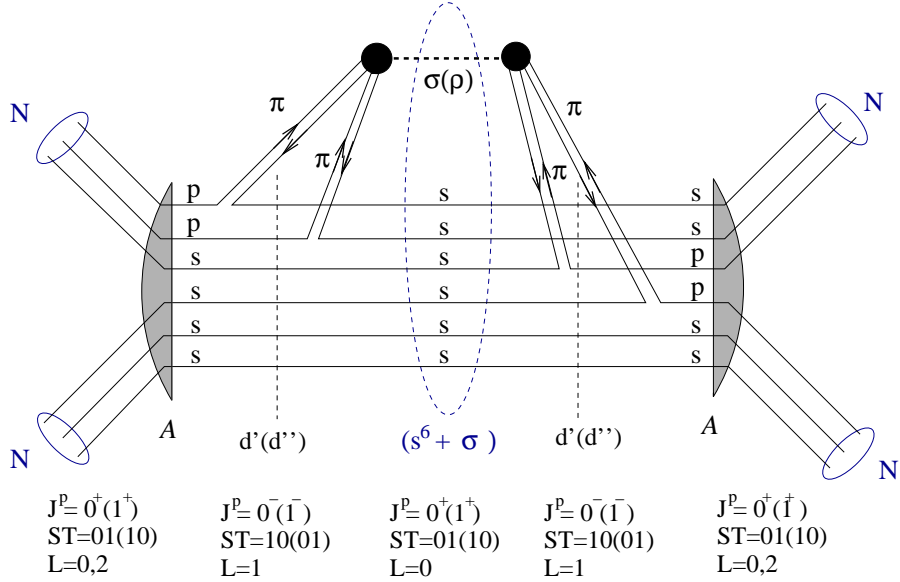}
\vspace{0cm}
\caption{\footnotesize The graph illustrates two sequential $\pi$-meson emission and absorptions via an intermediate $\sigma$-meson and the generation of a six-quark bag.}
\label{DBS_Sigma}
\end{figure}

\begin{figure}[t]
\begin{center}
\includegraphics[clip=true,width=0.6\textwidth]{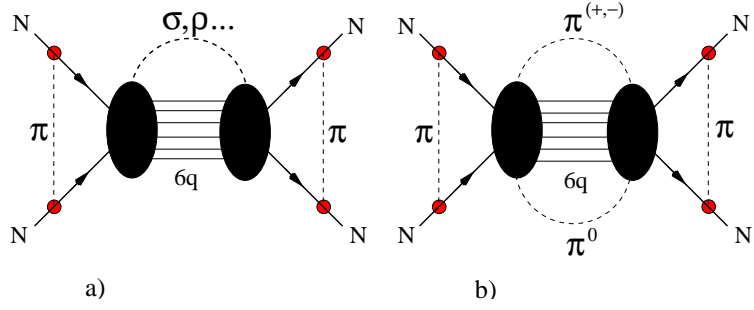}
\caption{\label{Figure1}
Two diagrams for $NN$ coupling to a six-quark bag with one $\sigma$- or
 $\rho$-meson (two correlated pions) or
  with two uncorrelated pions.}
\end{center}
\end{figure}

In conventional models the $NN$-potential appears when the mesonic degrees of
 freedom are eliminated explicitly from the nuclear wave function.  Contrary
 to this, in the new model the main part of the interaction appears when
 mesonic degrees of freedom are included explicitly. The key element of the
 new approach is a dense six-quark droplet surrounded by strong
 meson-fields, Fig.~\ref{DBS_Sigma}. 
 Naturally, such a bag has been called as Dressed Bag while the
 respective driving mechanism for the $NN$-force in the model is the
 Dressed-Bag-State (DBS) mechanism~\cite{2aa,4a}. Contrary to the conventional
 OBE-models, here the mesonic degrees of freedom are included in an
 {\it explicit} form. Thus the total wave function of the $NN$ system should
 be presented in form of eq.~(\ref{Fock2}).  The $NN$ wave function cam be
 generalised further
 with respect to the first version of the model of
 ref.~\cite{4a} by the important configurations which incorporate charged
 $2\pi$-clouds around the six-quark core (see Figs.~\ref{Figure1}a
 and~\ref{Figure1}b).

The channel couplings between different components are calculated from a
 quark-meson microscopic model~\cite{4a}. Typical diagrams representing the
 above couplings have a structure shown in Figs.~\ref{Figure1}a
 and~\ref{Figure1}b, where the $\sigma$-field represents the scalar-isoscalar
 channel, while the $s$-channel $2 \pi$-exchange (shown in
 Fig.~\ref{Figure1}b) is dominated by the isovector combination of two pions.

The $t$-channel (external) $\pi$- and $2 \pi$-exchanges are assumed to be
 usual Yukawa-type interactions with {\it soft} $\Lambda_{\pi NN}$ cut-offs,
 which represent the peripheral parts of the $NN$-force.  These soft cut-off
 form factors provide very natural separation between $t$-channel Yukawa-like
 (i.e. peripheral) $\pi$- and $2\pi$-exchanges and $s$-channel $\sigma$-,
 $\rho$- and $2\pi$-exchanges localised in the bag region, i.e. at $r < 1$~fm.
 The attractive interaction in the $NN$ channel in the DBS approach comes
 mainly from the $\sigma$-field (or $\sigma$-exchange in the $s$-channel)
 generated in the intermediate dense six-quark bag. The dominating dressing in
 even $NN$-partial waves ($S, D, \cdots$) is the $\sigma$-meson field while
 in odd partial waves this will come from the $\rho$-meson field.  It is also
 important that the initial and final $NN$-components (in $S$- and $D$-waves)
 are described by mixed symmetry six-quark states
 $|s^4p^2[42]_x;L=0,2,ST\rangle$ whereas the intermediate bag has a completely
 symmetric structure $|s^6 [6]_x; L=0, ST \rangle$; both six-quark components
 being orthogonal to each other.  Thus, the effective interaction in the $NN$
 channel will result after elimination of other Fock-components in the row
 (eq.~\ref{Fock2}) except $\Psi_{NN}$. The effective one-channel Hamiltonian
 for the $NN$-component takes the form~\cite{2aa,4a}:
\begin{equation}
\label{H_eff}
\mathcal{H}_{eff} = H_0 + V_{OPE} + V_{TPE} + 
\mathcal{V}_{NqN} + \lambda_{ort} \Gamma\ \ .
\end{equation}
Here $V_{OPE}$ and $V_{TPE}$ are the $\pi$- and $2\pi$-exchange potentials in
 the $t$-channel taken with the soft cut-off parameter $\Lambda_{\pi NN}
 \simeq 0.5$~GeV. $\mathcal{V}_{NqN}$ is the main part of the interaction at
 intermediate ranges, which is generated due to the intermediate
 DBS-production with dominating $\sigma$-meson dressing of the bag and
 includes inevitably an energy dependence which is the consequence of the
 elimination of some channels from the many-channel problem. The orthogonality
 pseudo-potential $\lambda_{ort} \Gamma$ with $\lambda_{ort} \to \infty$ (in
 practice the value of $\lambda$ is usually taken about $10^6$~MeV$\cdot$
 fm$^{-2}$) and
\begin{equation}
\Gamma = |0s \rangle \langle 0s |
\end{equation}
is the projection operator onto the $|0s \rangle$-state in the $NN$~channel,
 where the $\Gamma$ operator excludes the $|s^6[6]_x \rangle$ six-quark
 component from the initial $NN$~channel. As form factors $| 0s \rangle$ we
 use the h.o.  function with the common radius ~$r_0$:
\begin{equation}
|0s \rangle \equiv \frac{2}{\pi^{1/4} r_0^{3/2}} e^{-r^2/2r_0^2}.
\end{equation}

The $NN$-interaction in the fully symmetric six-quark state $[s^6]_x$ (which
 is undressed here) is strongly repulsive~\cite{8a} both in
 One-Goldstone-Boson (OGB)-exchange~\cite{9a,9aa} and in One-Gluon-Exchange
 (OGE)-models~\cite{10a,10aa,10aaa} (for $q-q$ interaction), and thus we
 replace this highly non-local $NN$-repulsion by the repulsive
 pseudo-potential $\lambda_{ort} \Gamma$ in a separable form. The forms for
 $V_{OPE}$ and $V_{TPE}$ adopted in this work are given in the Appendix to
 this paper.

The effective $NN$-interaction~$\mathcal{V}_{NqN}$ induced by the intermediate
  DBS generation has the following form:
\begin{equation}
\label{1S0}
\mathcal{V}_{NqN} = \lambda^{0}_{000} (E) |2s \rangle \langle 2s |,
\end{equation}
for the singlet $^1S_0$ channel. For the triplet $^3S_1 - {^3D_1}$
coupled channels we have
\begin{equation}
\label{3S1_3D1}
\mathcal{V}_{NqN} = 
\left(
\begin{array}{cc}
\lambda^{1}_{100} (E) |2s \rangle \langle 2s | & \,\,\,\,\,\,\,\,\,\,
\lambda^{1}_{101} (E) |2s \rangle \langle 2d | \\ \\
\lambda^{1}_{120} (E) |2d \rangle \langle 2s | & \,\,\,\,\,\,\,\,\,\,
\lambda^{1}_{122} (E) |2d \rangle \langle 2d |
\end{array}
\right)\ \ .
\end{equation}
 The bag radius is taken as $b=0.5$~fm and relates to
 h.o. radius~$r_0=\sqrt{2/3}~b$. 
 The standard normalised h.o.
 wave functions in the $NN$ relative motion
 $|2s\rangle$ and $|2d\rangle$ are formulated as follows:
\begin{equation}
|2s \rangle \equiv \sqrt{4 \pi} (\pi r_0^2)^{-3/4} \sqrt{\frac{3}{2}}
\left(1 - \frac{2 r^2}{3 r_0^2}\right) e^{-r^2/2r_0^2}\ \ ,
\end{equation}
\begin{equation}
|2d \rangle \equiv \sqrt{4 \pi}(\pi r_0^2)^{-3/4} \sqrt{\frac{4}{15}}
\left(\frac{r^2}{r_0^2}\right)  e^{-r^2/2r_0^2}\ \ .
\end{equation}

\begin{figure}[t]\centering
\includegraphics[scale=0.7, angle=0.]{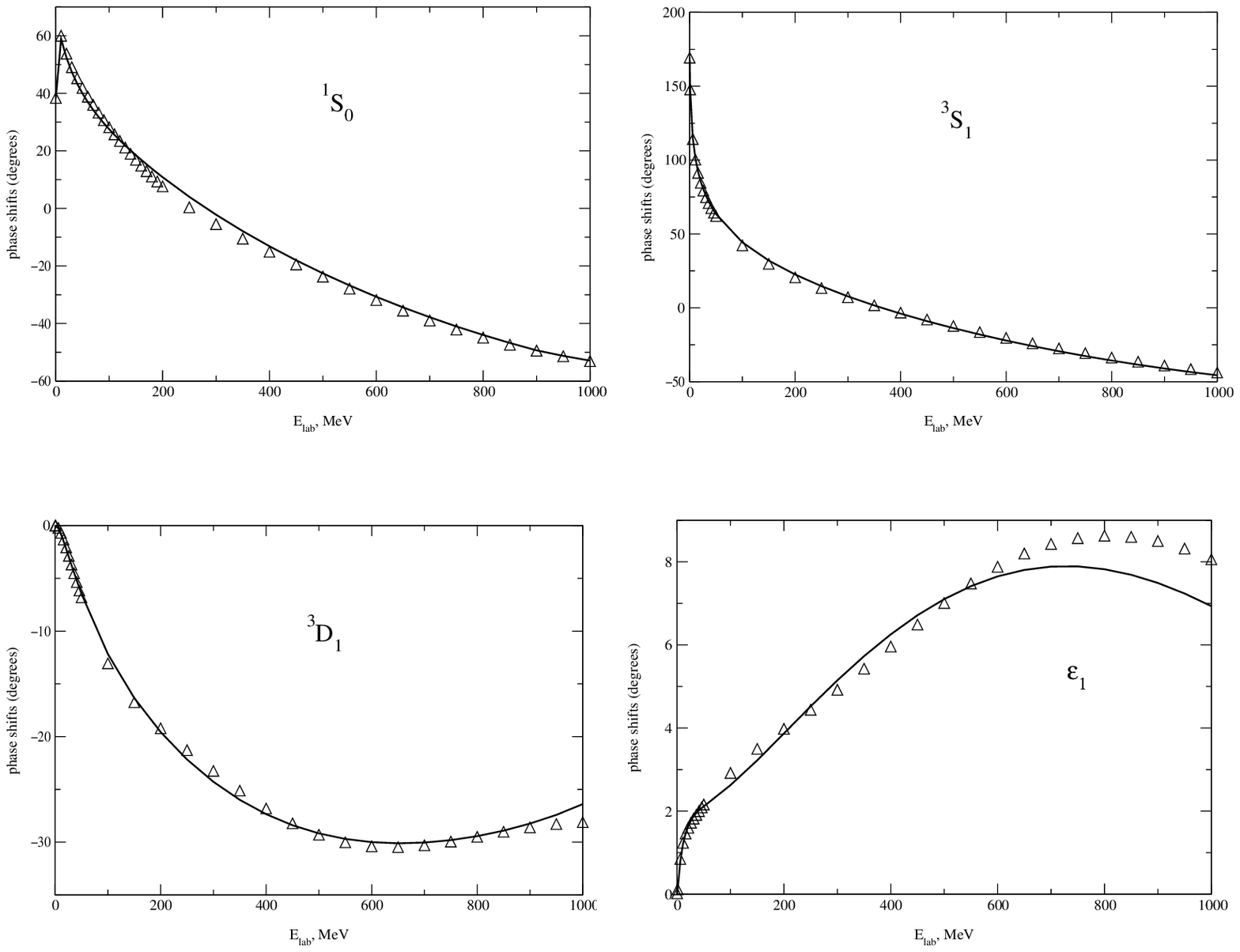}
\vspace{0cm}
\caption{\footnotesize $NN$-phase shifts in the DBS model up to 1 GeV}
\label{Fig3}
\end{figure}

The energy dependent coupling constants $\lambda_{SLL'}^J(E)$ in
 eqs.~(\ref{1S0}) and (\ref{3S1_3D1}) are calculated from microscopic $6q$-
 and $^3P_0$-quark pion coupling models~\cite{4a} and then are compared with
 those found from a fit to $NN$ phase shifts, Fig.~\ref{Fig3}, 
 from zero energy up to 1~GeV.  An
 impressive agreement for all relative values $\lambda_{SLL'^J}(0)$ was found.
 The energy dependence for $\lambda_{SLL'}^J(E)$ has been parameterised in
 ref.~\cite{4a} in the Pade-approximant form
\begin{equation}
\lambda^{J}_{SLL'}(E) = \lambda^{J}_{SLL'}(0)\ \frac{E_0 -a E}{E_0 - E}\ \ .
\end{equation}
However, the energy dependence $\lambda_{SLL'}^J(E)$ derived from the fit
 occurred to be somewhat stronger compared to the theoretical predictions. In
 the present work we use the values and the energy dependencies for
 $\lambda_{SLL'}^J(E)$ as tuned from the fit.

The parameters for $H_{eff}$ in eq.~(\ref{H_eff}), which provide a very
 good fit to $NN$-phase shifts from zero energy until as high as 1~GeV for
 $^3S_1-^3D_1$ and $^1S_0$-channels, are given in
 Table~\ref{MT_pot_parameters}.

\begin{table*}[h]
\caption{\label{MT_pot_parameters}
The potential parameters found by fitting the $NN$ scattering phase
  shifts up to 1~GeV (Ref.~\cite{4a}).}
\begin{center}
{\scriptsize
\begin{tabular}{c|ccccccccc}
\hline
\hline \\
  & $r_0$ & $r_2$ & $\lambda_{00}$ & $\lambda^{1}_{122}$ & 
$\lambda^{1}_{102}$  & $E_0$ & $ a$  & $V_{TPE}^0$ & $\beta$ \\ 
 & & & & & & & & & \\
 & fm & fm & MeV & MeV & MeV & MeV &  & MeV & fm$^{-2}$ \\ \\
\hline \\
$^3S_1-{^3D_1}$ & 0.41356 & 0.59423 & -328.55 & -15.65 & -44.06 & 693 & -0.05 
& -4.0573 & 0.5301 \\ \\
\hline \\
$^1S_0$ & 0.430 & - & -328.9 & - & - & 693 & -0.05 & -8.803 & 0.6441 \\ \\
\hline
\hline
\end{tabular}
}
\end{center}
\end{table*}

It should be stressed that the range and strength parameter values displayed
 in Table~\ref{MT_pot_parameters} are not only in good agreement with the
 values derived from the quark-meson microscopical calculations and empirical
 high-energy $NN$ phase shifts, but the same values of the parameters result
 in a {\it perfect} agreement for effective range parameters both in triplet
 and singlet channels; they also yield very good predictions for the deuteron
 structure (see below).

\section{The deuteron structure in the DBS model}
\label{sec:deut}

For this work on the deuteron we restrict ourselves to two components
of the full Fock column eq. (\ref{Fock2})
, which
can be represented by:
\begin{equation}\label{eq:deut2}
\Psi_d = \left(
\begin{array}{c}
\Psi_{NN} \\
\Psi_{6q + \sigma}
\end{array}
\right)
\end{equation}
The additional configurations generate new currents, which will be presented
in another paper, however they are of no relevance here.  According to
eq.~(\ref{eq:deut2}), the deuteron can be found in two different phases:
\begin{enumerate}
\item[1)]
   in the $NN$ cluster-like phase, in which its state is described in terms of
   the $NN$-relative motion variable ${\bf r}$, total spin $S_{NN} = S =1$,
   isospin $T_{NN} = T=0$, and the total angular momentum $J_{NN}= J =1$;
\item[2)]
   in the $6q+\sigma$ quark-meson phase, where one must use the dressed bag
   variables, i.e. the quark coordinates ${\bf r}_1$, ${\bf r}_2$, ...
   ${\bf r}_6$ together with their spins, colours, isospins, the total spin
   of the six-quark system ~$S_{6q}$, and also the momentum of the
   $\sigma$-meson~${\bf k}$ and the total angular momentum of the 
   bag $J_B = J = 1$.
\end{enumerate}
The first component describes the arrangement of the six quarks into nucleon
 clusters with the wave function given by the resonating group method (RGM) 
\footnote{The quark antisymmetrization operator~$\mathcal{A}$ is fully taken 
          into
          account in the calculation \cite{4a} for transitions between the 
          two phases
          $(NN)$ and $(6q + \sigma)$.  In the cluster-like
          $NN$ channel its effect reduces to a renormalisation of the
          $NN$~channel wave function with a constant factor of $\sqrt{10}$
          \cite{11a}.}:
\begin{equation}
\Psi_{NN} = \mathcal{A} \left[ \psi_N(123) \psi_N(456) \chi_{NN}({\bf r}) \right]
\end{equation}
where ${\bf r} \equiv ({\bf r}_1 + {\bf r}_2 + {\bf r}_3)/3 - 
({\bf r}_4 + {\bf r}_5 + {\bf r}_6)/3$. The $\psi_N(i,j,k)$ is the
nucleon wave function:
\begin{equation}
\psi(1,2,3) = \varphi(\vec{\xi}_1,\vec{\xi}_2) |~[1^3]_C S_{3q},
([21]_{CS}) T_{3q}: [1^3]CST \rangle
\end{equation} 
and the relative motion function $\chi_{NN}({\bf r})$ has the following
angular decomposition:
\begin{equation}
\chi_{NN}({\bf r}) = \frac{u(r)}{r} \mathcal{Y}_{1100}^{M_J 0}(\hat{\bf r})
+ \frac{w(r)}{r} \mathcal{Y}_{1120}^{M_J 0}(\hat{\bf r})
\end{equation}
with
\begin{equation}
\mathcal{Y}_{JSLT}^{M_J M_T}(\hat{\bf r}) =\sum_{M_l, M_s} (L M_L S M_S | J M_J)
Y_{L M_L} (\hat{\bf r}) |S M_S \rangle |T M_T \rangle 
\end{equation}
and $|S M_S \rangle$ and $|T M_T \rangle$ are the spin and isospin functions
for the two-nucleon deuteron state.

The DBS-component in simplest approximation (i.e. no additional pion loops are
 considered) can be described by the Yukawa-like wave function 
\begin{equation}
\Psi_{6q+\sigma} = \mathcal{N} \varphi_{6q} \cdot \chi(r_{\sigma})
\end{equation}
where $\mathcal{N}$ is the normalising factor. We use the abbreviations
$|\varphi_{6q} \rangle = |s^6[6]_x; L=0,ST \rangle $ and
$\ds \chi(r_{\sigma}) \sim \frac{\exp(-\mu_\sigma r_{\sigma})}{\mu_\sigma r_{\sigma}}$,
 with 
$r_{\sigma}$ being the  distance of the $\sigma$-meson from the six-quark
core. The $\sigma$ mass is not fixed strictly in this model because it
represents a quasi-particle defining the quantum numbers of the
scalar-isoscalar channel. It depends slightly upon the particular system and the
type of the process under investigation.
 Thus, $\mu_\sigma \simeq 400$~MeV \cite{PDG} is a sensible choice.

\begin{table*}[h]
\caption{\label{Deut_prop}
 Deuteron properties in different models.}
\begin{center}
{\scriptsize
\begin{tabular}{c|lllllll}
\hline
\hline \\
\mbox{Model} & $E_d$(MeV) & $P_D(\%)$ & $r_m$(fm) & $Q_d$(fm$^2$) &
$\mu_d(\mu_N)$ & $A_S$(fm$^{-1/2}$) & $D/S$ \\ \\
\hline \\
RSC       & 2.22461  & 6.47 & 1.957 & 0.280 & 0.8429 & 0.8776 & 0.0262 \\ \\
AV18      & 2.2245   & 5.76 & 1.967 & 0.270 & 0.8521  & 0.8850 & 0.0256   \\ \\
Moscow 99$^{a)}$ & 2.22452  & 5.52 & 1.966 & 0.272 & 0.8483 & 0.8844 & 0.0255 \\ \\
Bonn 2001 & 2.22458 & 4.85 & 1.966 & 0.270 & 0.8521 & 0.8846 & 0.0256 \\ \\
DBS model & 2.22454  & 5.42 & 2.004 & 0.286 & 0.8489 & 0.9031 & 0.0259 \\ 
($NN$ component only)& & & & & & & \\ \\
DBS model & 2.22454  & 5.22 & 1.972
& 0.275 & 0.8548 & 0.8864 & 0.0264  \\ 
$NN + 6q (3.66\%)$ & & & & & & & \\ \\
\hline \\
Experiment & 2.2245475(9)$^{b)}$ &  & 1.971(3)$^{c)}$
& 0.2859(3)$^{d)}$ & 0.857406(1)$^{e)}$ & 0.8846(44)$^{f)}$ & 0.0264$^{g)}$ \\ \\
\hline
\hline 
\end{tabular}
}
\end{center}
\hspace{1.cm} $~^{a}$) Ref.~\cite{11a};
\hspace{0.3cm} $^{b}$) Ref.~\cite{Ed};
\hspace{0.3cm} $^{c}$) Ref.~\cite{b};
\hspace{0.3cm} $^{d}$) Refs.~\cite{c1,c2};
\hspace{0.3cm} $^{e}$) Ref.~\cite{mud};
\hspace{0.3cm} $^{f}$) Ref.~\cite{d};
\hspace{0.3cm} $^{g}$) Ref.~\cite{12a} 
\end{table*}

Having chosen all the parameters of the effective Hamiltonian
 $\mathcal{H}_{eff}$ in the $NN$~channel, the deuteron two-phase structure can
 be derived with no additional parameters.  In Table~\ref{Deut_prop} the main
 static properties of the deuteron as emerging from the above $NN$-model are
 compared to experiment and to predictions of other modern $NN$-force models.

The weight of the DBS-component in the deuteron wave function, as calculated
 from the above model, is 3.66\%\footnote{The contribution of the DBS 
 to r.m.s. radius of matter in deuteron
is proportional to the weight of the DBS-state. However due to much wider radial distribution in nucleon component of the deuteron, as compared to the DBS,
the real correction to the r.m.s. radius of matter in the deuteron is rather 
small $(\sim 0.03$ fm); see Table~\ref{Deut_prop}, sixth and seventh lines in the third column.}.
          {We emphasise here that the six-quark component weight in this model
          is a direct consequence of the six-quark model parameters
          established either theoretically or from the fit to $NN$ phase
          shifts. And thus this weight cannot be varied or fitted to other
          data.}
This weight looks to be quite significant and it is within the limits found
with other hybrid models. A reasonable variation of the radius
          parameter~$r_0$ for the $6q$-core in connection with the
          readjustment of the strength~$\lambda$ does not result in any
          significant change of the weight for the DBS component in the
          deuteron. This holds also for other observables. 
 The level of agreement between theoretical
predictions of this model and the most accurate experimental data for deuteron
observables is not worse than for the best modern $NN$-potentials
 of conventional
type.  Our results for the $np \to d \gamma$ process presented in the paper can
be fully compared with those of other modern $NN$-models.

\begin{figure}[t]
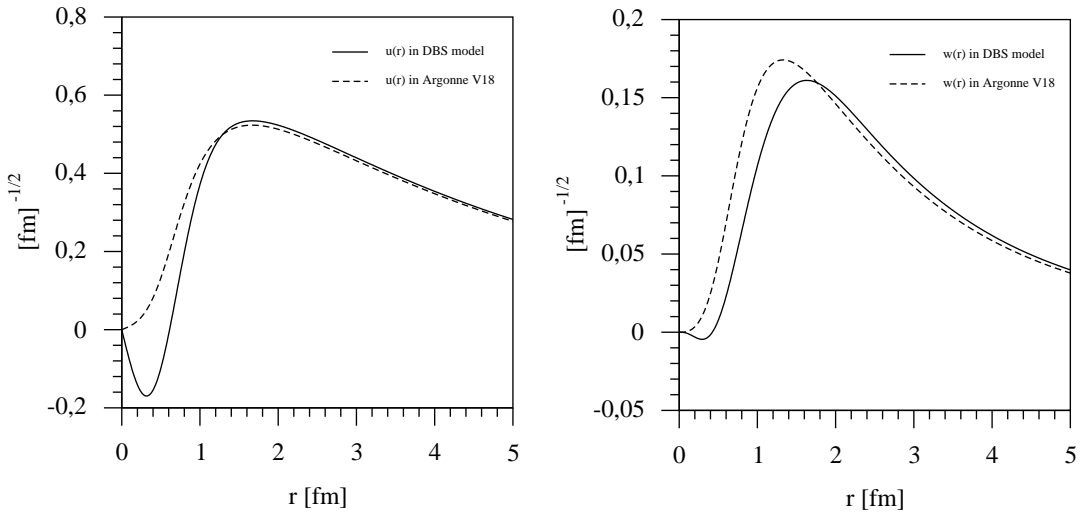

\parbox[t]{7.5cm}{
\begin{center}
\includegraphics[clip=true,scale=0.55,angle=0.]{Figure1a.eps}
\end{center}
}
\parbox[t]{6.5cm}{
\begin{center}
\includegraphics[clip=true,scale=0.55,angle=0.]{Figure1b.eps}
\end{center}
}
\caption{\label{DWF}
 Deuteron wave functions $u(r)$ and $w(r)$ calculated in the six-quark
 dressed bag model.}
\end{figure}
It is also of interest to compare the form of the deuteron wave functions
 predicted by our model and traditional ones; see Fig.~\ref{DWF} for both the
 $S$- and the $D$-wave components.  It is evident from Fig.~\ref{DWF} that the
 DBS model generates a deuteron $S$-wave function with a node at short
 distances ($r \sim 0.5 \div 0.6$~fm).  This node is a consequence of the
 requirement of orthogonality between all $S$- (and $P$-) wave functions in
 the $NN$ channel and the quark part of the $DBS$-state, projected onto the
 $NN$ channel. There is also a rather small short-range loop in the deuteron
 $D$-wave component of DBS-wave function.

Thus, two-component wave functions are present in our approach, both in the
 initial and in the final (i.e. deuteron) channel. It should be stressed, that
 both components have a different structure and both will contribute to the
 $np \to d \gamma$ transition amplitude, and the contribution of the
 DBS-component will by no means be negligible. Our main concern in this paper
 is the comparison between predictions of our model and the conventional
 $NN$-potential models, with the inclusion of both the nucleon currents and
 MEC-contributions. Because the conventional models have no other
 contributions we postpone the study of the new currents inside the DBS-model
 to future work. We present in this paper the comparison with traditional
 OBE-models in the $NN$-sector and the conventional MEC-operators.

The wave functions of the $np$ continuum and the deuteron are obtained by
 solving the Schr\"odinger equation with $\mathcal{H}_{eff}$ from
 eq.~(\ref{H_eff}).  In the present calculation the same potential model is
 used to determine both the initial scattering state as well as the deuteron
 wave function.

\section{General transition electromagnetic currents}
\label{sec:transcurr}

From a quantitative point of view, the relationship between the nuclear wave
 function, the $NN$-potential and \mbox{e.-m.} operators is expressed via a
 continuity equation, which ensures that the total current is conserved
 (global gauge invariance).  The e.-m. current operator
 ${\bf j}$ of a nuclear many-body system is related to the Hamiltonian of the
 system by the continuity equation
\begin{equation}
\label{Cont_eqn}
{\vec \nabla} \cdot {\bf j} + i [H, \rho] = 0,
\end{equation}
where $\rho$ is the charge operator of the system. So that once the total
 system Hamiltonian is known, eq.~(\ref{Cont_eqn}) gives the current operator
 $\vec{j}$ for the meson-exchange-current (MEC) calculations.

The contribution of MEC to e.-m. processes on nuclei constitutes an important
 manifestation of meson degrees of freedom for charged mesons mediating the
 strong interactions between nucleons in nuclei. They participate in the
 interaction of an external e.-m. field with a nucleus and result
 formally in the appearance of nuclear currents in the form of two- or many-body
 current operators. So they are intimately linked to the underlying
 $NN$-interaction. However, for a given $NN$-potential there does not exists an
 a-priori way of constructing the appropriate MEC-operator, unless the
 $NN$-potential is derived from an underlying more fundamental
 field-theoretical framework with explicit incorporation of subnuclear degrees
 of freedom. Although the existence of exchange currents associated with an
 $NN$-potential has been acknowledged for a long time, the early realistic
 potentials, being phenomenological to a large extent, prevented thus the
 consistent consideration of such exchange currents
\footnote{One can remark, however, that the minimal account of mesonic degrees
  of freedom in. e.-m. transitions has been taken for a long time through the
  use of the Siegert theorem.}.

The situation is complicated by the fact that the e.-m. current operator
 contains irreducible two-body exchange current components, which through the
 continuity equation depend on the potential model.  Therefore, consistent
 gauge invariant calculations require in general, that the same potential
 model must be used to generate both the wave functions and the exchange
 current operators.

The full hadronic current operator has the following matrix form
in the DBS model:
\begin{eqnarray}
\hat{j}_{\mu}(x) = 
\left(
\begin{array}{cc}
\hat{j}^{NN}_{\mu}(x) \,\,\,\,\,\, & \hat{j}^{NN \to DBS}_{\mu}(x) \\
\hat{j}^{DBS \to NN}_{\mu}(x) \,\,\,\,\,\, & \hat{j}^{DBS}_{\mu}(x)
\end{array}
\right) 
\end{eqnarray}
where the diagonal currents
\begin{eqnarray}
\label{Cl_currens}
\hat{j}^{NN}_{\mu}(x) &=& \hat{j}^{IA}_{\mu}(x) + \hat{j}^{MEC}_{\mu}(x)
+ \hat{j}^{IC}_{\mu}(x) \\
\label{DBS_current}
\hat{j}^{DBS}_{\mu}(x) &=& 
\hat{j}^{cloud}_{\mu}(x) + \hat{j}^{6q}_{\mu}(x)
\end{eqnarray}
correspond to the interaction of an external e.-m. field with the $NN$ and the
 Dressed Bag, respectively. 
The third term in eq. (\ref{Cl_currens}) corresponds to the isobar
current (IC) associated with the intermediate $\Delta(1232)$ excitation.
The transition (non-diagonal) currents
 $\hat{j}^{NN \to DBS}_{\mu}$ or $\hat{j}^{DBS \to NN}_{\mu}$ describe the
 interphase transitions induced by the external e.-m. field.

We consider here only that part of the hadronic e.-m. current
 operator $\hat{j}^{NN}_{\mu}$ given by eq.~(\ref{Cl_currens}), which is
 associated with cluster-like nucleon-nucleon configurations.

\subsection{Impulse approximation}
\label{ssec:ia}

The simplest description of nuclei is based on a nonrelativistic many-body
 theory of interacting nucleons. Within this framework, the nuclear
 e.-m. operators are expressed in terms of those associated with the
 individual protons and neutrons --- the so-called impulse approximation (IA).
 In this approximation the one-body nuclear e.-m. $\rho({\vec q})$ and ${\vec
 j}({\vec q})$ operators are obtained from the covariant single-nucleon
 current~\cite{Carlson98}:
\begin{equation}
\label{4_current}
j^{\mu} = \bar u ({\vec p}') \left[ F_1({\mathcal Q}^2) \gamma^{\mu} +
F_2({\mathcal Q}^2) \frac{i \sigma^{\mu \nu} q_{\nu}}{2 M_N} \right] u ({\vec p}),
\end{equation}
where ${\vec p}$ and ${\vec p}'$ are the initial and final momenta,
 respectively, of a nucleon of mass $m_N$, and $F_1({\mathcal Q}^2)$ and
 $F_2({\mathcal Q}^2)$ are its Dirac and Pauli form factors taken as a
 function of the four-momentum transfer:
\begin{equation}
{\mathcal Q}^2 = - q_{\mu} q^{\mu}, \,\,\,\,\, 
q_{\mu} = p_{\mu}' - p_{\mu}.
\end{equation}
The current~$j_{\mu}$ is expanded in powers of $1/m_N$ and, including terms up
 to an order of $1/m_N$, the nuclear current operator can be written as a sum
 of one-body current operators which takes the form~\cite{Schiav91}:
\begin{eqnarray}
{\vec j}({\vec q}) = \sum_i \frac{1}{4 M_N} 
\left[ {\mathcal G}^S_E({\mathcal Q}^2) + {\mathcal G}^V_E({\mathcal Q}^2) 
\tau_{z,i} \right] ({\vec p}_{i}' + {\vec p}_{i}) e^{i {\vec q}{\vec r}_i}  
\nonumber \\
+ \frac{i}{4 M_N}
\left[{\mathcal G}^S_M({\mathcal Q}^2) + 
{\mathcal G}^V_M({\mathcal Q}^2) \tau_{z,i} \right]
[\vec \sigma_i \times {\vec q}] e^{i {\vec q}{\vec r}_i}
\nonumber \\
\end{eqnarray}
where the first term is the so-called convection current and the second is
 the magnetisation current.

\begin{figure*}[t]
\hspace{-7cm}
\begin{picture}(100,50)
\thicklines
\multiput(84,21.5)(2,2){14}{\oval(2,2)[lt]}
\multiput(82,21.5)(2,2){14}{\oval(2,2)[rb]}
\put(61,13){\line(-1,1){15}}
\put(61,13){\line(-1,-1){15}}
\put(61,13){\circle*{10}}
\put(105,13){\line(1,1){15}}
\put(105,13){\line(1,-1){15}}
\put(93,13){\circle*{10}}
\put(105,13){\circle*{10}}
\put(73,13){\circle*{10}}
\put(83,20.5){\circle*{1}}
\put(83,13){\oval(18,15)}
{\thinlines
\multiput(50,2)(0,2.84){8}{\line(0,1){2}}
\multiput(116,2)(0,2.84){8}{\line(0,1){2}}
\put(67,14){\oval(10,10)[t]}
\put(99,14){\oval(10,10)[t]}
\put(63,14.25){\line(1,0){10}}
\put(63,13.75){\line(1,0){10}}
\put(63,13.25){\line(1,0){10}}
\put(63,12.75){\line(1,0){10}}
\put(63,12.25){\line(1,0){10}}
\put(63,11.73){\line(1,0){10}}
\put(95,14.25){\line(1,0){10}}
\put(95,13.75){\line(1,0){10}}
\put(95,13.25){\line(1,0){10}}
\put(95,12.75){\line(1,0){10}}
\put(95,12.25){\line(1,0){10}}
\put(95,11.73){\line(1,0){10}}
\multiput(78,-2)(0,2){25}{\line(0,1){1}}
\multiput(88,-2)(0,2){25}{\line(0,1){1}}
}
\put(94,37){{$\gamma$}}
\put(98,20.5){{$\sigma$}}
\put(66,20.5){{$\sigma$}}
\put(65,7){{$6q$}}
\put(97,7){{$6q$}}
\put(43,22){{$N$}}
\put(119,22){{$N$}}
\put(43,1){{$N$}}
\put(119,1){{$N$}}
\put(118,12){{$\pi$}}
\put(46,12){{$\pi$}}
\put(67,42){{$\mathcal{A}$}}
\put(82,42){{$\mathcal{B}$}}
\put(95,42){{$\mathcal{C}$}}
\end{picture}
\vspace{1cm}
\caption{\label{fig:3}
One-body e.-m. current in IA which is constructed in the DBS model on the
 basis of the diagrams presented in Fig.~\ref{Figure1}. Region~$\mathcal{A}$
 corresponds to the generation of the initial $NN$ scattering wave function:
 in the region~$\mathcal{B}$, the emission of a $\gamma$-quantum by a single nucleon
 takes place and leads to the $M1$ transition; region~$\mathcal{C}$ generates
 the final deuteron wave function.
}
\end{figure*}
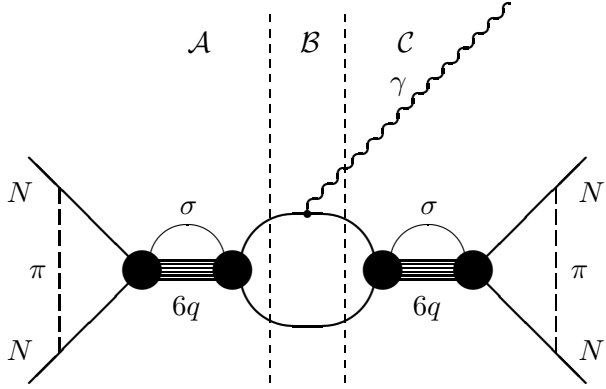

The Sachs form factors ${\mathcal G}_E$ and ${\mathcal G}_M$ are related to
 the Dirac and Pauli form factors in eq.~(\ref{4_current}) 
via~\cite{Carlson98}:
\begin{equation}
{\mathcal G}_E({\mathcal Q}^2) = F_1({\mathcal Q}^2) 
- \frac{{\mathcal Q}^2}{4 M^2_N} F_2({\mathcal Q}^2),
\end{equation}
\begin{equation}
{\mathcal G}_M({\mathcal Q}^2) = F_1({\mathcal Q}^2) + F_2({\mathcal Q}^2)
\end{equation}
and are normalised so that
\begin{equation}
{\mathcal G}^S_E({\mathcal Q}^2 = 0) = {\mathcal G}^V_E({\mathcal Q}^2 = 0) =
1, ~~
{\mathcal G}^S_M({\mathcal Q}^2 = 0) = \mu_p + \mu_n = 0.880~\mu_N, ~~ 
\end{equation}
\begin{equation}
{\mathcal G}^V_M({\mathcal Q}^2 = 0) = \mu_p - \mu_n = 4.706~\mu_N
\end{equation}
where $\mu_p$ and $\mu_n$ are the magnetic moments of the proton and neutron
 in terms of the nuclear magneton $\mu_N$.  The superscripts $S$ and $V$ of
 the Sachs form factors ${\mathcal G}_E$ and ${\mathcal G}_M$ denote isoscalar
 and isovector combinations of the proton and neutron electric and magnetic
 form factors, respectively.  Then the transition matrix elements take the
 form
\begin{equation}
\mathcal{M} = 
\langle \chi^{^{3}S_{1}-^{3}D_1}_d\ | {\vec j}({\vec q}) \cdot \vec{\epsilon}_{\lambda} |
\, \chi^{^{1}S_{0}}_{np} \rangle
\end{equation}
where
\begin{equation}
| \chi^{^{3}S_{1}}_{d}({\vec r}) \rangle = \frac{1}{\sqrt{4 \pi}} \frac{u(r)}{r} | 1 M_S \rangle_S  | 0 0  \rangle_T  
\end{equation}
and
\begin{eqnarray}
| \chi^{^{3}D_{1}}_{d}({\vec r}) \rangle  = 
\frac{w(r)}{r} \sum_{M_L M_S} 
(2 M_L 1 M_S | 1 M_J) Y_{2M_L}(\hat{\bf r}) |1 M_S \rangle_S | 0 0  \rangle_T  
\end{eqnarray}
are $^3S_1$ and $^3D_1$-deuteron wave function components, respectively.

For the initial $np$ $^{1}S_0$ continuum state we use a similar form
\begin{eqnarray}
| \chi^{^{1}S_{0}}_{np} \rangle =
\frac{1}{\sqrt{4 \pi}} \frac{z(r)}{r} | 0 0  \rangle_S | 1 0  \rangle_T   
\end{eqnarray}

The direct calculation for the impulse approximation~(IA) results in the
following expressions for the amplitudes~$\mathcal{M}$ 
\begin{eqnarray}
\mathcal{M}_{^1S_0 \to ^3S_1}=  
G_{^1S_0 \to {^3S_1}}^{IA} 
\,\, \sqrt{\frac{8 \pi}{3}}\, 
\sum_{k} (1 M_J 1 k \ 1 \lambda) Y_{1k}(\hat{\bf q}) 
\end{eqnarray}
\begin{eqnarray}
\mathcal{M}_{^1S_0 \to ^3D_1}
=  
- \, G_{^1S_0 \to ^3D_1}^{IA}  
\,\, \sqrt{\frac{8 \pi}{3}}\, 
\sum_{k} (1 M_J 1 k \ 1 \lambda) Y_{1k}(\hat{\bf q}) 
\end{eqnarray}
where the functions $G_{^1S_0 \to {^3S_1}}^{IA}$ and 
$G_{^1S_0 \to {^3D_1}}^{IA}$ are taken as follows:
\begin{eqnarray}
G_{^1S_0 \to {^3S_1}}^{IA} = 
{\mathcal G}^V_M({\mathcal Q}^2 = 0)
\ \frac{q}{2 M_N}  \,
\left[ \int_{0}^{\infty} u(r) z(r) j_0 (\frac{qr}{2})  d\, r \right]
\end{eqnarray}
\begin{eqnarray}
G_{^1S_0 \to {^3D_1}}^{IA} = 
{\mathcal G}^V_M({\mathcal Q}^2 = 0) 
\ \frac{q}{2M_N}  \,
\left[ \frac{1}{\sqrt 2} \int_{0}^{\infty} w(r) z(r) j_2 (\frac{qr}{2})
  d\, r \right]
\end{eqnarray}

\noindent and with
\begin{equation}
G_{^1S_0 \to d}^{IA} =\left[ G_{^1S_0 \to {^3S_1}}^{IA} - 
G_{^1S_0 \to {^3D_1}}^{IA} \right] 
\end{equation}

\begin{eqnarray}
\mathcal{M}_{IA} =
G_{^1S_0 \to d}^{IA}
\,\, \sqrt{\frac{8 \pi}{3}}\,   
\sum_{k} (1 M_J 1 k \ 1 \lambda) Y_{1k}(\hat{\bf q}) 
\end{eqnarray}

The IA cross section is thus proportional to the quantity
 $\left[G_{^1S_0 \to d}^{IA}\right]^2$ as one can 
 neglect the states other than $^1S_0$ in the entrance channel
 because of its low energy:
\begin{eqnarray}
\label{IA_res}
G_{^1S_0 \to d}^{IA} =
{\mathcal G}^V_M({\mathcal Q}^2 = 0) 
\ \frac{q}{2M_N} 
\int_{0}^{\infty} \left[ u(r) z(r) j_0 (\frac{qr}{2})  
- \frac{1}{\sqrt 2} w(r) z(r) j_2 (\frac{qr}{2}) \right]d\, r
\end{eqnarray}

The result of the destructive interference between $G^{^1S_0 \to
  {^3S_1}}_{IA}$ and $G^{^1S_0 \to {^3D_1}}_{IA}$ is a sharp minimum in the
  cross section for the $e + d \to e + n + p$ reaction which has not been seen
  experimentally.  Notice that at high momenta the transition to the $D$-state
  of the deuteron is as important as the transition to the $S$-state. The
  total matrix element changes its sign at about $q^2_c \approx 12 fm^{-2}$.

Although the IA has been utilised frequently in investigations of weak and
 e.-m. interactions in nuclei, it does not preserve all the salient features
 characterising the ``elementary'' interaction with free nucleons. For
 instance, even though the e.-m. current for on-shell nucleons is conserved as
 implied by gauge invariance, the {\it nuclear} or hadronic e.-m. currents
 generated by the conventional IA are not conserved in general and are
 inconsistent with gauge invariance~\cite{Hwange80}. To restore current
 conservation in conventional IA, we should incorporate also MEC and take
 seriously the argument that only the sum of all the contributions shall be
 conserved. In other words, we have to take seriously the two basic
 ingredients of the IA, viz.: (1) The interaction with the whole nucleus can
 be approximated by a simple sum of the ``elementary'' interactions with the
 constituent nucleons, and (2) each elementary interaction with the
 constituent nucleon takes place {\it instantaneously}.  By
 ``instantaneously'' it is understood, that the elementary interaction takes
 place in region~$\mathcal{B}$ of Fig.~\ref{fig:3} and thus is not affected by
 the presence of the initial and final state interactions. Even if the
 off-shell effects arising from the binding of constituent nucleons by the
 nuclear potential can be neglected, the masses for the initial and final
 states of the nucleon cannot always be treated 
 as the same
 because of the instantaneous absorption of the four-momentum
 transfer. Current conservation is already violated in this elementary
 interaction if, in spite of such difference in the effective masses, we adopt
 as our basic input the form factors for free nucleons. To guarantee that the
 resultant impulse approximation is in accord with current conservation, we
 have to attempt to restore current conservation for each elementary
 interaction by generalising the on-shell form factors to include the
 off-shell effects due to, in particular, the instantaneous absorption of the
 four-momentum transfer~\cite{Hwang80_2}.

\subsection{Pion-exchange currents with the soft cut-off parameters}
\label{ssec:pioncurr}

It is important to note here that both the pion-exchange interaction and the
 model independent pion-exchange current operator may be derived as
 nonrelativistic limits of the corresponding operators obtained from a
 relativistic chiral symmetric Lagrangian model for the meson-nucleon
 system. The two-body currents generated from the meson degrees of freedom can
 be classified according to their time (charge) and space (current)
 components, and also to their isospin structure (isoscalar or
 isovector). These contributions are of different nature, the
 isovector current is of the same $1/M$ order as the one-body current, while
 all other transitions appear at the relativistic level and should compete
 with relativistic corrections to the impulse approximation.

The diagrams for MEC of longest range are presented in Figs.~\ref{Seagull_DBS}
 and~\ref{Mesonic_DBS}. All transition operators representing these 
 exchange-current processes have
 isovector character, and do therefore lead to transition from the deuteron
 state to the $^1S_0$ state and not to the $\alpha$-state
 $({^3S_1}~+~{^3D_1})$. There is,
 of course, no limit to the number of possible contributing exchange-current
 processes, but only those of pionic range are really significant, since
 short-range heavy-meson exchange processes are suppressed by the strong
 short-range repulsion in the nucleon-nucleon interaction~\cite{Hockert73}.
 It means that a isoscalar counterpart of the 
    isovector currents, due to 
    $\rho \pi \gamma$ and $\omega \pi \gamma$ contributions
    contribute only a little to the isovector 
    transition    
    under consideration and thus have been neglected in the present study.
    Besides that,
    the class of model-dependent currents  
    between different mesons, i.e. $\rho \pi \gamma$ ($\omega \pi
    \gamma$), are purely transverse and are of short range, which is due to
    the large $\rho$- ($\omega$) mass.  Therefore, their contribution at low
    momentum transfer should be very small~\cite{Carlson98}.  
    It should be emphasized also that the contribution related to these
    operators are rather sensitive to the values of cut-off
    parameters. For the soft cut-off parameters used in the DBS model
    ($\Lambda_{\pi} \sim 0.7$ GeV) these contributions will be even less
    important than in the traditional OBEP models where $\Lambda_{\pi} \sim 1.2$
    GeV is usually employed.

\subsubsection{The Seagull term}
\label{sssec:sea}

\begin{figure*}[t]
\hspace{-7cm}
\begin{picture}(100,45)
\thicklines
\multiput(84,21.5)(2,2){14}{\oval(2,2)[lt]}
\multiput(82,21.5)(2,2){14}{\oval(2,2)[rb]}
\put(61,13){\line(-1,1){15}}
\put(61,13){\line(-1,-1){15}}
\put(61,13){\circle*{10}}
\put(105,13){\line(1,1){15}}
\put(105,13){\line(1,-1){15}}
\put(93,13){\circle*{10}}
\put(105,13){\circle*{10}}
\put(73,13){\circle*{10}}
\put(83,20.5){\circle*{1}}
\put(83,13){\oval(18,15)}
{\thinlines
\multiput(50,2)(0,2.84){8}{\line(0,1){2}}
\multiput(116,2)(0,2.84){8}{\line(0,1){2}}
\multiput(83,20.5)(0,-2.6){6}{\line(0,-1){2}}
\put(67,14){\oval(10,10)[t]}
\put(99,14){\oval(10,10)[t]}
\put(63,14.25){\line(1,0){10}}
\put(63,13.75){\line(1,0){10}}
\put(63,13.25){\line(1,0){10}}
\put(63,12.75){\line(1,0){10}}
\put(63,12.25){\line(1,0){10}}
\put(63,11.73){\line(1,0){10}}
\put(95,14.25){\line(1,0){10}}
\put(95,13.75){\line(1,0){10}}
\put(95,13.25){\line(1,0){10}}
\put(95,12.75){\line(1,0){10}}
\put(95,12.25){\line(1,0){10}}
\put(95,11.73){\line(1,0){10}}
\multiput(78,-2)(0,2){25}{\line(0,1){1}}
\multiput(88,-2)(0,2){25}{\line(0,1){1}}
}
\put(94,37){{$\gamma$}}
\put(98,20.5){{$\sigma$}}
\put(66,20.5){{$\sigma$}}
\put(65,7){{$6q$}}
\put(97,7){{$6q$}}
\put(43,22){{$N$}}
\put(119,22){{$N$}}
\put(43,1){{$N$}}
\put(119,1){{$N$}}
\put(118,12){{$\pi$}}
\put(46,12){{$\pi$}}
\put(84,12){{$\pi$}}
\put(67,42){{$\mathcal{A}$}}
\put(82,42){{$\mathcal{B}$}}
\put(95,42){{$\mathcal{C}$}}
\end{picture}
\vspace{1cm}
\caption{\label{Seagull_DBS}
The seagull term contribution in the DBS model. 
Notations for the regions are the same as in
 Fig.~\ref{fig:3} for the one-body current.
}
\end{figure*}
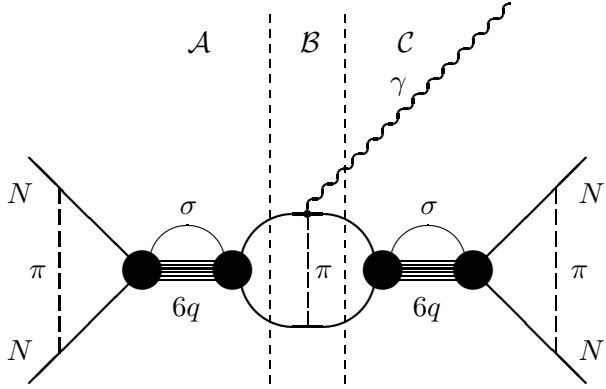

The most important MEC process is due to the seagull current 
operator~${\bf J}^{sea}_{\pi}$ which is a two-body current operator (see
 Fig.~\ref{Seagull_DBS}):
\begin{eqnarray}
\label{Seagull_cur}
{\bf J}^{sea}_{\pi} ({\vec k}_1, {\vec k}_2) = 
\mbox{i} \,
\frac{f^2_{\pi NN}}{m^2_{\pi}} (\vec{\tau} _1 \times \vec{\tau} _2 )^z
\left[ \frac{ \vec{\sigma} _2 ( \vec{\sigma} _1 \cdot {\vec k}_1 )}
{m^2_{\pi} + {\vec k}^2_1} - 
\frac{ \vec{\sigma} _1 ( \vec{\sigma} _2 \cdot {\vec k}_2 )}
{m^2_{\pi} + {\vec k}^2_2} \right]
\end{eqnarray} 

The configuration-space expression~\cite{Carlson98} may be obtained from 
 the respective Fourier transform of ${\bf J}^{sea}_{\pi}({\vec k}_1,{\vec k}_2)$:
\begin{eqnarray}
{\bf J}^{sea}_{\pi} ({\vec q}) = 
\int d^3 \, {\vec x} \,\, e^{i {\vec q} \cdot {\vec x}}
\int \frac{d^3 \, {\vec k}_{1}}{(2 \pi)^3}
\frac{d^3 \, {\vec k}_{2}}{(2 \pi)^3}
\,\, e^{i {\vec k}_1 \cdot ({\vec r}_1 - {\vec x}) }
\ e^{i {\vec k}_2 \cdot ({\vec r}_2 - {\vec x}) }
\ {\bf J}^{sea}_{\pi} ({\vec k}_1, {\vec k}_2)
\end{eqnarray}
which results in:
\begin{eqnarray}
\label{Seagull_conf}
{\bf J}^{sea}_{\pi} ({\vec q}) = -
\frac{f^2_{\pi NN}}{m^2_{\pi}} (\vec{\tau} _1 \times \vec{\tau} _2 )^z
(m_{\pi} + \frac{1}{r}) \frac{ e^{- m_{\pi} r}}{4 \pi r} 
\left\{ 
\vec{\sigma} _2 ( \vec{\sigma} _1 \cdot \hat{\bf r}  ) 
e^{{i} {\vec q} \cdot {\vec r}_2}
+ 
\vec{\sigma} _1 ( \vec{\sigma} _2 \cdot \hat{\bf r}  )
e^{{i} {\vec q} \cdot {\vec r}_1} 
\right\} 
\end{eqnarray}
Thus the transition matrix element is of the form:
\begin{equation}
\mathcal{M}_{^1S_0 \to d}^{sea} =  
\langle{\psi_d} | {\bf J}^{sea}_{\pi}({\vec q}) \cdot 
\vec{\epsilon}_{\lambda} | \psi_{pn} \rangle
\end{equation}
It can be naturally decomposed in two terms:
\begin{eqnarray}
\mathcal{M}_{^1S_0 \to d}^{sea} = 
\left[ G_{^1S_0 \to ^3S_1}^{sea} + G_{^1S_0 \to ^3D_1}^{sea} \right]
\times
\sqrt{\frac{8 \pi}{3}}\,
\sum_{i} (1 M_J 1 i |  1 \lambda) Y_{1i}(\hat{\bf q}) 
\end{eqnarray}
where the functions $G_{^1S_0 \to {^3S_1}}^{sea}$ and 
$G_{^1S_0 \to {^3D_1}}^{sea}$ are defined as:
\begin{eqnarray}
\label{Seagull_cur_conf}
G_{^1S_0 \to ^3S_1}^{sea} = 
\left( \frac{f^2_{\pi NN}}{4 \pi} \right) \frac{4}{m^2_{\pi}}
\left[ \int_{0}^{\infty} u(r) z(r) j_1 (\frac{qr}{2})  
(1 + m_{\pi}r) \frac{ e^{- m_{\pi} r}}{r^2} d\, r \right] 
\end{eqnarray}

\begin{eqnarray}
\label{Seagull_cur_confd}
G_{^1S_0 \to ^3D_1}^{sea} = 
\left( \frac{f^2_{\pi NN}}{4 \pi} \right) \frac{4}{m^2_{\pi}}
\left[ \frac{1}{\sqrt 2} \int_{0}^{\infty} w(r) z(r) j_1 (\frac{qr}{2})  
(1 +m_{\pi}r ) \frac{ e^{- m_{\pi} r}}{r^2} d\, r \right] 
\end{eqnarray}

\subsubsection{The Pion-in-flight term} 
\label{sssec:pif}

\begin{figure*}[h]
\hspace{-7.cm}
\begin{picture}(100,50)
\thicklines
\multiput(85,14)(2,2){14}{\oval(2,2)[lt]}
\multiput(83,14)(2,2){14}{\oval(2,2)[rb]}
\put(61,13){\line(-1,1){15}}
\put(61,13){\line(-1,-1){15}}
\put(61,13){\circle*{10}}
\put(105,13){\line(1,1){15}}
\put(105,13){\line(1,-1){15}}
\put(93,13){\circle*{10}}
\put(105,13){\circle*{10}}
\put(73,13){\circle*{10}}
\put(83,13){\circle*{1}}
\put(83,13){\oval(18,15)}
{\thinlines
\multiput(50,2)(0,2.84){8}{\line(0,1){2}}
\multiput(116,2)(0,2.84){8}{\line(0,1){2}}
\multiput(83,20.5)(0,-2.6){6}{\line(0,-1){2}}
\put(67,14){\oval(10,10)[t]}
\put(99,14){\oval(10,10)[t]}
\put(63,14.25){\line(1,0){10}}
\put(63,13.75){\line(1,0){10}}
\put(63,13.25){\line(1,0){10}}
\put(63,12.75){\line(1,0){10}}
\put(63,12.25){\line(1,0){10}}
\put(63,11.73){\line(1,0){10}}
\put(95,14.25){\line(1,0){10}}
\put(95,13.75){\line(1,0){10}}
\put(95,13.25){\line(1,0){10}}
\put(95,12.75){\line(1,0){10}}
\put(95,12.25){\line(1,0){10}}
\put(95,11.73){\line(1,0){10}}
\multiput(78,-2)(0,2){25}{\line(0,1){1}}
\multiput(88,-2)(0,2){25}{\line(0,1){1}}
}
\put(97,33){{$\gamma$}}
\put(98,20.5){{$\sigma$}}
\put(66,20.5){{$\sigma$}}
\put(65,7){{$6q$}}
\put(97,7){{$6q$}}
\put(43,22){{$N$}}
\put(119,22){{$N$}}
\put(43,1){{$N$}}
\put(119,1){{$N$}}
\put(118,12){{$\pi$}}
\put(46,12){{$\pi$}}
\put(79,15){{$\pi$}}
\put(79,9){{$\pi$}}
\put(67,42){{$\mathcal{A}$}}
\put(82,42){{$\mathcal{B}$}}
\put(95,42){{$\mathcal{C}$}}
\end{picture}
\vspace{1cm}
\caption{\label{Mesonic_DBS}
The pion-in-flight term contribution in the DBS model.  
Notations for the regions are the same as in
 Fig.~\ref{fig:3} for the one-body e.-m. current.
}
\end{figure*}
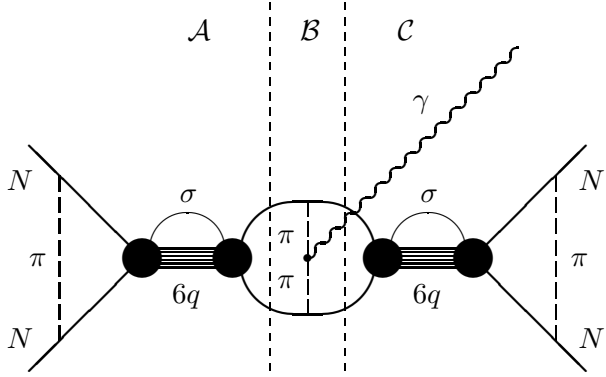

Among all the conventional diagrams, the pion-in-flight term is the most
 difficult one to calculate, since it does involve two-pion propagators (see
 Fig.~\ref{Mesonic_DBS}).
This diagram represents the direct coupling of the photon to the exchanged
 pion. It corresponds to virtual pion photoproduction on a single nucleon by the
 pion-pole amplitude, followed by absorption on the second one. The result of
 the calculation for this pion-in-flight (pif) term is:
\begin{eqnarray}
\label{Mesonic_cur}
{\bf J}^{pif}_{\pi} ({\vec k}_1, {\vec k}_2) =  
\mbox{i} \,
\frac{f^2_{\pi NN}}{m^2_{\pi}} (\vec{\tau} _1 \times \vec{\tau} _2 )^z
({\vec k}_1 - {\vec k}_2)
 \frac{ ( \vec{\sigma} _1 \cdot {\vec k}_1 )
(\vec{\sigma} _2 \cdot {\vec k}_2) }
{(m^2_{\pi} + {\vec k}^2_1)(m^2_{\pi} + {\vec k}^2_2)} 
\end{eqnarray} 
And the transition matrix elements can be written as follows:
\begin{equation}
\mathcal{M}_{^1S_0 \to d}^{pif} =  
\langle{\psi_d} | {\bf J}^{pif}_{\pi}({\vec q}) \cdot 
\vec{\epsilon}_{\lambda} | \psi_{pn} \rangle
\end{equation}
consisting of two components:
\begin{eqnarray}
\mathcal{M}_{^1S_0 \to d}^{pif} =  
\left[ G_{^1S_0 \to ^3S_1}^{pif} + G_{^1S_0 \to ^3D_1}^{pif} \right]
\sqrt{\frac{8 \pi}{3}}\,
\sum_{i} (1 M_J 1 i |  1 \lambda) Y_{1i}(\hat{\bf q}) 
\end{eqnarray}
where the functions $G_{^1S_0 \to {^3S_1}}^{pif}$ and 
$G_{^1S_0 \to {^3D_1}}^{pif}$ take the forms:
\begin{eqnarray}
\label{Mesonic_cur_conf}
G_{^1S_0 \to ^3S_1}^{pif} = 
\left( \frac{f^2_{\pi NN}}{4 \pi} \right) \frac{2}{3}
\frac{q}{m^2_{\pi}}
\,\,  
\left[ \int_{0}^{\infty} 
\left\{ \left( \frac{d^2}{dr^2} + \frac{2}{r} \frac{d}{dr} \right)
t_0 
\right. \right. \hspace{5cm}
\nonumber \\
\left. \left. 
+ \left( \frac{d^2}{dr^2} 
+ \frac{5}{r} \frac{d}{dr} + \frac{3}{r^2} 
\right) t_2 \right\} 
u(r) z(r)  d\, r \right] \hspace{0.5cm}
\end{eqnarray}

\begin{eqnarray}
\label{Mesonic_cur_confd}
G_{^1S_0 \to ^3D_1}^{pif} = 
\left( \frac{f^2_{\pi NN}}{4 \pi} \right) \frac{2}{3}
\frac{q}{m^2_{\pi}}
\,\,  
\left[ \int_{0}^{\infty} 
\left\{ \left( \frac{d^2}{dr^2} - \frac{1}{r} \frac{d}{dr} \right)
t_0 
\right. \right. \hspace{5cm}
\nonumber \\
\left. \left. 
+ \left( \frac{d^2}{dr^2} + \frac{2}{r} \frac{d}{dr} - \frac{6}{r^2} 
\right) t_2 \right\} 
\frac{1}{\sqrt 2} w(r) z(r)  d\, r \right] \hspace{0.5cm} 
\end{eqnarray}
where
\begin{equation}
\label{t_l}
t_l = \int_0^1  \frac{e^{- \alpha r}}{\alpha} j_l (\frac{\eta q r}{2}) d \eta,
~~~~~~~ l = 0,2
\end{equation}
and
\begin{equation}
\alpha = \sqrt{m_{\pi}^2 + \frac{1}{4} q^2 (1-\eta^2)}.
\end{equation}
In the long-wave limit (limit for small $q r$) the Bessel functions
$\ds j_l (\frac{\eta q r}{2})$ of the eq. (\ref{t_l}) 
can be reduced according to:
\begin{equation}
j_l(z) \stackrel{z \to 0}{\longrightarrow} \frac{z^l}{(2 l + 1)!!}, 
~~~~~~~ (2 l + 1)!! = 1 \cdot 3 \cdot 5 \cdots \cdot (2 l +1),
\end{equation}
so that the eqs. (\ref{Mesonic_cur_conf}) and (\ref{Mesonic_cur_confd}) 
result in:
\begin{equation}
G_{^1S_0 \to ^3S_1}^{pif} = 
\left( \frac{f^2_{\pi NN}}{4 \pi} \right) \frac{2}{3}
\frac{q}{m^2_{\pi}}
\,\,  
\left[ \int_{0}^{\infty} 
\left\{ (m_{\pi}r - 2) \frac{e^{-m_{\pi}r}}{r}  \right\} 
u(r) z(r)  d\, r \right] 
\end{equation}
\begin{equation}
G_{^1S_0 \to ^3D_1}^{pif} = 
\left( \frac{f^2_{\pi NN}}{4 \pi} \right) \frac{2}{3}
\frac{q}{m^2_{\pi}}
\,\,  
\left[ \frac{1}{\sqrt 2} \int_{0}^{\infty} 
\left\{ (m_{\pi}r + 1) \frac{e^{-m_{\pi}r}}{r}  \right\} 
 w(r) z(r)  d\, r \right] 
\end{equation}

\subsubsection{The $\Delta(1232)$-isobar current term}
\label{sssec:delta}

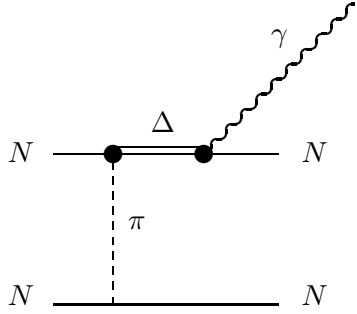
\begin{figure*}[h]
\hspace{-8cm}
\begin{picture}(100,50)
{\thicklines
\multiput(96,31)(2,2){10}{\oval(2,2)[rb]}
\multiput(98,31)(2,2){10}{\oval(2,2)[lt]}
\put(96,30){\circle*{2.5}}
\put(84,30){\circle*{2.5}}
}
{\thinlines
\put(76,30){\line(1,0){30}}
\put(76,10){\line(1,0){30}}
\put(84,31){\line(1,0){12}}
\multiput(84,10)(0,2){10}{\line(0,1){1}}
}
\put(89,33){{$\Delta$}}
\put(86,20){{$\pi$}}
\put(70,29){{$N$}}
\put(109,29){{$N$}}
\put(70,10){{$N$}}
\put(109,10){{$N$}}
\put(105,45){{$\gamma$}}
\end{picture}
\caption{\label{Delta_cur}
The $\Delta$-isobar current diagram.}
\end{figure*}

The process corresponding to the virtual excitation of a $\Delta(1232)$
 followed by the $\Delta N \to NN$ transition interaction
 (Fig.~\ref{Delta_cur}) is described by the $\Delta$-current~${\bf
 J}^{\Delta}_{\pi}$.  There are many approaches to evaluate this effect. We
 assume $\Delta$-dominance, and we shall use here the static quark model and
 the sharp resonance approximation.

In the static quark model the exchange current due to the $\Delta$
resonance is \cite{Hockert73}:
\begin{eqnarray}
{\bf J}^{\Delta}_{\pi} = \mbox{i} \frac{ g^2_{\pi NN} \mathcal{G}_{M}^{V}
(\mathcal{Q}^2 =0)}{25 M_N^3 (M_{\Delta}-M_N)}
\left\{4 \tau_3^1 \frac{(\vec k _1 \times \vec q )
(\vec \sigma ^1 \cdot \vec k _1)}{m_{\pi}^2 + \vec k _1^2} 
+
4 \tau_3^2 \frac{(\vec k _2 \times \vec q )
(\vec \sigma ^2 \cdot \vec k _2)}{m_{\pi}^2 + \vec k _2^2}
\right. \hspace{3cm}
\nonumber \\
-
\left.
(\vec \tau ^1 \times \vec \tau ^2)_3 
\left[ \frac{(\vec \sigma ^1 \times \vec k _2) \times \vec q 
(\vec \sigma ^2 \times \vec k _2)}{m_{\pi}^2 + \vec k _2^2}
-
\frac{(\vec \sigma ^2 \times \vec k _1) \times \vec q 
(\vec \sigma ^1 \times \vec k _2)}{m_{\pi}^2 + \vec k _1^2}
\right]
\right\}~~~~
\end{eqnarray}
here $\vec k _1$ and $\vec k _2$ are the fractional momentum
transfers imparted to the first and second nucleons, and
$g_{\pi NN}$ is the pseudoscalar $\pi NN$ coupling constant, which is
related to the pseudovector $\pi NN$ coupling constant $f_{\pi N N}$ as:
$$ \frac{g_{\pi NN}}{2 M_N} = \frac{f_{\pi N N}}{m_{\pi}}
~~~~~~\mathrm{and}
~~~~~~ \frac{f^2_{\pi N N}}{4 \pi} = 0.075\ \ .$$
The transition matrix element for the $\Delta$-exchange current is:
\begin{equation}
\mathcal{M}_{^1S_0 \to d}^{\Delta} =  
\langle{\psi_d} | {\bf J}^{\Delta}_{\pi}({\vec q}) \cdot 
\vec{\epsilon}_{\lambda} | \psi_{pn} \rangle .
\end{equation}
consisting of two contributions:
\begin{eqnarray}
\mathcal{M}_{^1S_0 \to ^3S_1}^{\Delta} = 0
\end{eqnarray}
and
\begin{eqnarray}
\mathcal{M}_{^1S_0 \to ^3D_1}^{\Delta} = 
G_{^1S_0 \to ^3D_1}^{\Delta} 
\,\, \sqrt{\frac{8 \pi}{3}}\,
\sum_{i} (1 M_J 1 i |  1 \lambda) Y_{1i}(\hat{\bf q}),
\end{eqnarray}
where $G_{^1S_0 \to ^3D_1}^{\Delta}$ in the long-wave limit  
has the form:
\begin{eqnarray}
G_{^1S_0 \to ^3D_1}^{\Delta} = 
\mathcal{G}_{M}^{V} (\mathcal{Q}^2 =0)
\ \frac{g^2_{\pi NN}}
{4 \pi} 
\frac{q}{M_{\Delta} - M_N}
\left(\frac{m_{\pi}}{M_N}\right)^3
\hspace{3.2cm}
\nonumber \\
\times
\frac{4}{25}
\left[ \frac{2}{3} \int_{0}^{\infty} w(r) z(r)   
\left(1 + \frac{3}{m_{\pi}r} + \frac{3}{(m_{\pi}r)^2} \right) \frac{ e^{- m_{\pi} r}}{m_{\pi} r} d\, r \right].
\end{eqnarray}

\subsubsection{Effect of short range correlations}
\label{sssec:src}

One important feature of the $\pi$-exchange currents which has not been much
 emphasised in previous studies of the radiative neutron capture is the
 sensitivity of these exchange-current effects to the $\pi NN$ form factor.
From a general point of view, the $\pi NN$ form factor originates due to the
 modification of the bare $\pi$-exchange current at short
 distances~\cite{Leidemann83}, where the finite size of the nucleon (and the
 pion) together with the vertex renormalisation associated with very
 short-range multi-pion exchange processes (the coupling of the pion to
 heavier systems such as $\rho$ or $3\pi$) can be of some
 importance~\cite{Mathiot82}.  The former corrections are roughly related to
 the confinement radius of the quarks in the nucleon.

It is rather difficult to calculate the form factors exactly, as they involve
 many non-perturbative contributions
\footnote{One rather popular modern approach to the reliable determination of
          the electroweak and strong $\pi NN$ form factors is based on well
          known QCD sum rules.}.
For practical applications they are usually parametrised in momentum space by
 the monopole form, the mass scale $\Lambda$ being a parameter chosen in order
 to be consistent with the experiment. All the consistent estimates given up
 to date lead to a cut-off parameter $\Lambda_{\pi NN}$ lying around $0.5 \div
 0.8$~GeV/c.  In the same way as the e.-m. form factor in MEC operators has
 been related to the longitudinal part of the charge operator from the
 continuity equation, the hadronic form factors in these operators should be
 connected to the choice of a particular $NN$-potential.

One takes
\begin{equation}
F_{\pi NN}(q) = \frac{\Lambda^2_{\pi} - m^2_{\pi}}{\Lambda^2_{\pi} + {\vec q}^2}
\end{equation}
at each $\pi NN$ vertex. The values of the cut-off masses are, of course, to
 be related to the cut-offs used in the construction of the $NN$-potential.

\begin{figure*}[t]
\hspace{-7cm}
\begin{picture}(100,30)
{\thicklines
\put(76,20){\line(1,0){7}}
\put(97,20){\line(-1,0){7}}
}
{\thinlines
\put(76,10){\line(0,1){20}}
\put(97,10){\line(0,1){20}}
\multiput(84,20)(2,0){3}{\line(1,0){1}}
}
\put(92,22){{$\Pi$}}
\put(78,22){{$\Pi$}}
\put(85,22){{$\pi$}}
\put(69,27){{$N$}}
\put(100,27){{$N$}}
\put(69,10){{$N$}}
\put(100,10){{$N$}}
\end{picture}
\vspace{-0.5cm}
\caption{\label{Modif_OPEP}
The one-pion-exchange potential (OPEP) modified by the hadronic form
  factor of monopole type at each vertex.}
\end{figure*}
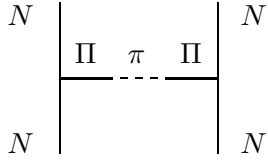

\begin{figure*}[t]
\begin{center}
\hspace{-8cm}
\begin{picture}(100,40)
{\thicklines
\put(26,20){\line(1,0){7}}
\put(47,20){\line(-1,0){7}}
\multiput(26,21)(-2,2){10}{\oval(2,2)[lb]}
\multiput(24,21)(-2,2){10}{\oval(2,2)[rt]}
\put(26,20){\circle*{1.5}}
}
{\thinlines
\put(26,10){\line(0,1){20}}
\put(47,10){\line(0,1){20}}
\multiput(34,20)(2,0){3}{\line(1,0){1}}
}
\put(41,16){{$\Lambda_{\pi}$}}
\put(28,16){{$\Lambda_{\pi}$}}
\put(35,16){{$\pi$}}
\put(20,27){{$N$}}
\put(49,27){{$N$}}
\put(20,10){{$N$}}
\put(49,10){{$N$}}
\put(10,29){{$\gamma$}}
\put(35,2){{$(a)$}}
{\thicklines
\put(66,20){\line(1,0){7}}
\put(87,20){\line(-1,0){7}}
\multiput(70,23)(0,4){4}{\oval(2,2)[l]}
\multiput(70,21)(0,4){4}{\oval(2,2)[r]}
\put(70,20){\circle*{1}}
}
{\thinlines
\put(66,10){\line(0,1){20}}
\put(87,10){\line(0,1){20}}
\multiput(74,20)(2,0){3}{\line(1,0){1}}
}
\put(81,16){{$\Lambda_{\pi}$}}
\put(68,16){{$\Lambda_{\pi}$}}
\put(75,16){{$\pi$}}
\put(60,27){{$N$}}
\put(89,27){{$N$}}
\put(60,10){{$N$}}
\put(89,10){{$N$}}
\put(73,30){{$\gamma$}}
\put(75,2){{$(b)$}}
{\thicklines
\put(106,20){\line(1,0){7}}
\put(127,20){\line(-1,0){7}}
\multiput(117,23)(0,4){4}{\oval(2,2)[l]}
\multiput(117,21)(0,4){4}{\oval(2,2)[r]}
\put(117,20){\circle*{1}}
}
{\thinlines
\put(106,10){\line(0,1){20}}
\put(127,10){\line(0,1){20}}
\multiput(114,20)(2,0){3}{\line(1,0){1}}
}
\put(121,16){{$\Lambda_{\pi}$}}
\put(108,16){{$\Lambda_{\pi}$}}
\put(115,16){{$\pi$}}
\put(100,27){{$N$}}
\put(129,27){{$N$}}
\put(100,10){{$N$}}
\put(129,10){{$N$}}
\put(120,30){{$\gamma$}}
\put(110,2){{$(c)$}}
{\thicklines
\put(146,20){\line(1,0){7}}
\put(167,20){\line(-1,0){7}}
\multiput(164,23)(0,4){4}{\oval(2,2)[l]}
\multiput(164,21)(0,4){4}{\oval(2,2)[r]}
\put(164,20){\circle*{1}}
}
{\thinlines
\put(146,10){\line(0,1){20}}
\put(167,10){\line(0,1){20}}
\multiput(154,20)(2,0){3}{\line(1,0){1}}
}
\put(161,16){{$\Lambda_{\pi}$}}
\put(148,16){{$\Lambda_{\pi}$}}
\put(155,16){{$\pi$}}
\put(140,27){{$N$}}
\put(169,27){{$N$}}
\put(140,10){{$N$}}
\put(169,10){{$N$}}
\put(159,30){{$\gamma$}}
\put(155,2){{$(d)$}}
\end{picture}
\caption{\label{Modif_Mesonic}
The seagull (a) and the pion-in-flight (b,c,d) diagrams modified by hadronic 
         form factors of monopole form. The diagrams (b) and (d) are required
         in order to ensure gauge invariance}
\end{center}
\end{figure*}
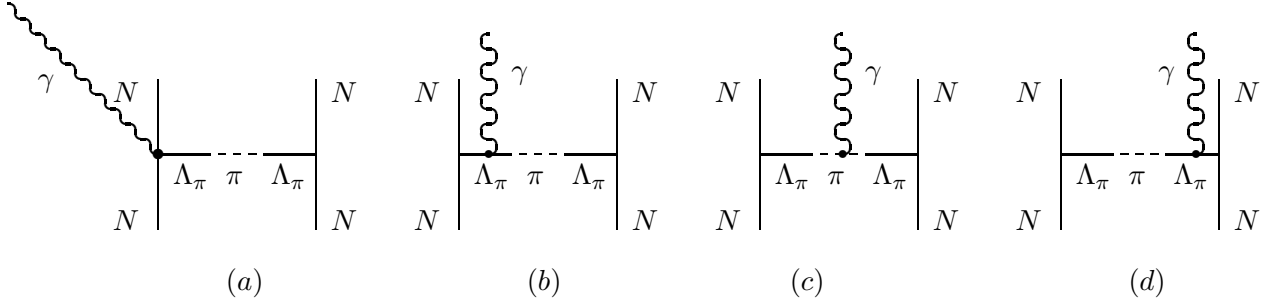

One important point in deriving the expressions for MEC with the hadronic form
 factors is to insure gauge invariance, or in other words, the current
 conservation. It has been shown a long time ago that current conservation is
 insured, if one takes a $\pi$-exchange model for the $NN$-potential together
 with exchange currents (seagull and mesonic) including an appropriate
 e.-m. form factor of nucleons and pions but without any meson-nucleon form
 factor. The next step is the verification whether insertion of the hadronic
 form factor at each meson-nucleon vertex satisfies this requirement too. The
 $\pi$-exchange potential is modified by a monopole form factor at each vertex
 in the following way:
\begin{equation}
V_{ij}^{OPE}({\vec q}) = - \frac{f_{\pi NN}^2}{m^2_{\pi}} 
\left( \frac{\Lambda^2_{\pi} - m^2_{\pi}}{\Lambda^2_{\pi} + {\vec q}^2}
\right)^2 
\frac{({\vec \sigma}_i \cdot {\vec q})({\vec \sigma}_j \cdot {\vec q})}{{\vec
    q}^2 + m^2_{\pi}} ({\vec \tau}_i \cdot {\vec \tau}_j),
\end{equation} 
Apart from the normalisation factor $\ds (\Lambda^2_{\pi} - m^2_{\pi})^2$ this
 modified potential can be visualised by the diagram shown in
 Fig.~\ref{Modif_OPEP}, where $\Pi$ represents a particle of mass
 $\Lambda_{\pi}$ with the same quantum numbers as the $\pi$-meson (the
 short-range vertex renormalisation is represented by an exchange of a virtual
 heavy particle, of mass $\Lambda_{\pi}$).  By minimal substitution on this
 diagram, it is then straightforward to construct the corresponding exchange
 currents, modified by hadronic form factors and satisfying gauge
 invariance. These are depicted in Fig.~\ref{Modif_Mesonic}a (modified seagull
 term) and Fig.~\ref{Modif_Mesonic}b,c,d (modified mesonic term).

The seagull current operator is just modified by adding the corresponding
 factor $F^2_{\pi NN}(q)$ in the numerator of the expression given in
 eq.~(\ref{Seagull_cur}):
\begin{eqnarray}
{\bf J}^{sea}_{\pi} ({\vec k}_1, {\vec k}_2) = \mbox{i} \,
\frac{f^2_{\pi NN}}{m^2_{\pi}} (\vec{\tau} _1 \times \vec{\tau} _2 )^z 
\left[ \frac{ \vec{\sigma} _2 ( \vec{\sigma} _1 \cdot {\vec k}_1 )}
{m^2_{\pi} + {\vec k}^2_1} 
\frac{\Lambda^2_{\pi} - m_{\pi}^2}{\Lambda^2_{\pi} + {\vec k}^2_1} - 
\frac{ \vec{\sigma} _1 ( \vec{\sigma} _2 \cdot {\vec k}_2 )}
{m^2_{\pi} + {\vec k}^2_2}
\frac{\Lambda^2_{\pi} - m_{\pi}^2}{\Lambda^2_{\pi} + {\vec k}^2_2}
 \right] \hspace{0.5cm}
\end{eqnarray}
The leading isovector $\pi$-exchange pion-in-flight current operator given in
 eq.~\ref{Mesonic_cur} is thus modified:
\begin{eqnarray}
{\bf J}^{pif}_{\pi} ({\vec k}_1, {\vec k}_2) = \mbox{i} \,
\frac{f^2_{\pi NN}}{m^2_{\pi}} (\vec{\tau} _1 \times \vec{\tau} _2 )^z
({\vec k}_1 - {\vec k}_2) 
(\Lambda_{\pi}^2 - m_{\pi}^2)
( \vec{\sigma} _1 \cdot {\vec k}_1 )  ( \vec{\sigma} _2 \cdot {\vec k}_2 ) 
\hspace{2cm}
\nonumber \\
\times
\left\{ \frac{1}{\Lambda^2_{\pi} + {\vec k}^2_1} 
\right.
\left[ 
\frac{1}{(\Lambda^2_{\pi} + {\vec k}^2_2)(m^2_{\pi} + {\vec k}^2_2)} 
+ \frac{1}{(m^2_{\pi} + {\vec k}^2_1)(m^2_{\pi} + {\vec k}^2_2)} 
\right.
\nonumber \\
\left. \left.
+ \frac{1}{(m^2_{\pi} + {\vec k}^2_1)(\Lambda^2_{\pi} + {\vec k}^2_1)} 
\right] \frac{1}{\Lambda_{\pi}^2 + {\vec k}^2_2}
\right\} 
\end{eqnarray}

The main difference from the usual derivation of exchange currents without
 form factors is the appearance of the two new contributions, shown in
 Fig.~\ref{Modif_Mesonic}a and Fig.~\ref{Modif_Mesonic}c, which arise in the
 mesonic term.  After the inclusion of the hadronic form factor, the
 exchange-current operators in eqs.~(\ref{Seagull_cur_conf}) and
 (\ref{Mesonic_cur_conf}) are modified in the following way:
\begin{eqnarray}
\label{Curr_sea_FF}
\frac{e^{-m_{\pi}r}}{r^2} (1 + m_{\pi}r) \Longrightarrow 
\frac{e^{-m_{\pi}r}}{r^2} (1 + m_{\pi}r) 
- \frac{e^{-\Lambda_{\pi}r}}{r^2} (1 + \Lambda_{\pi}r) -
\frac{\Lambda_{\pi}^2 - m_{\pi}^2}{2} e^{-\Lambda_{\pi} r}, 
\end{eqnarray}
\begin{eqnarray}
\label{Curr_mes_FF}
\frac{e^{- \alpha r}}{\alpha} 
\Longrightarrow 
\frac{e^{- \alpha r}}{\alpha}  
- \frac{e^{- \lambda r}}{\lambda} -
\frac{( \Lambda_{\pi}^2 - m_{\pi}^2)}{2} \frac{(1+ \lambda r)}{\lambda^3}
e^{- \lambda r} 
\end{eqnarray}
where
\begin{equation}
\lambda = \sqrt{\Lambda_{\pi}^2 + \frac{1}{4} q^2 (1 - \eta^2)}.
\end{equation}

Thus, eqs.~(\ref{Seagull_cur_conf}), (\ref{Seagull_cur_confd}),
           (\ref{Mesonic_cur_conf}), (\ref{Mesonic_cur_confd}),
           (\ref{Curr_sea_FF}) and (\ref{Curr_mes_FF})
 define the exchange currents within the conventional model to be used in this
 work. This model has the attractive feature that it satisfies current
 conservation consistent with the one-pion exchange potential. The effect of
 the short-range hadronic form factor to the $\Delta$-isobar contribution in
 the $NN$ wave function was investigated by transition potential
 methods~\cite{Mathiot84} with $\pi$-exchange contributions. For the
 calculations we employed a potential proposed in ref.~\cite{Durso77}.

\section{Cross sections for the thermal neutron capture}
\label{sec:xsec}

The thermal $np$ capture is a fundamental process for our purpose because it
 involves two opposite features, long-distance behaviour and short-range
 processes, where the latter are not completely understood. Clearly, a better
 determination of the asymptotic behaviour of the theoretical $NN$ wave
 function is needed before a quantitative comparison with experiment can be
 made for the thermal $np$ capture cross section.

For thermal neutrons we can use a nonrelativistic kinematic regime, and we may
 even ignore the deuteron recoil energy. Then the $\gamma$-ray emitted in the
 capture process will have an energy essentially equal to the deuteron binding
 energy $\epsilon_d$ \cite{Adler72}. In this kinematic regime the differential
 cross section has the form

\begin{equation}
d \, \sigma_{fi} = \frac{\delta (P_{f} - P_{i})}{\left| {\bf v}_n -{\bf v}_p
 \right|} 
 \frac{e^2}{(2 \pi)^2} \frac{1}{2 \omega_{\gamma}}
\left| \mathcal{M}_{fi} \right|^2 d^3 {\vec p}_d d^3 {\vec q}_{\gamma} 
\end{equation}
The integrals over ${\bf p}_d$ and $|{\bf q}|$ in the laboratory system where
$|{\bf v}_p| =0$, are easy to take~\cite{Adler72} if we ignore the deuteron
recoil and set $q = \omega = \epsilon_d$, i.e. the deuteron binding energy.
So, these integrations results in: 
\begin{equation}
\label{Therm_cross_sec}
\frac{d \, \sigma_{fi}}{d \, \Omega_{\gamma}} = 
\frac{\omega_{\gamma}}{\left| {\bf v}_n \right|}
 \frac{e^2}{2(2 \pi)^2} 
\left| \mathcal{M}_{fi} \right|^2 
\end{equation}
Since we neglect the deuteron recoil momentum, there is no angular dependence
 in the differential cross section~$d \sigma_{fi}/ d \Omega_{\gamma}$ due to
 phase space. Therefore, all the angular dependence is due to the dynamics of
 the capture process. 

The total transition amplitude takes the form:
\begin{equation}
\mathcal{M}_{{^1S_{0}} \to d} = \mathcal{M}^{IA}_{{^1S_{0}} \to d} +
\mathcal{M}^{sea}_{{^1S_{0}} \to d} + \mathcal{M}^{pif}_{{^1S_{0}} \to d} +
\mathcal{M}_{^1S_0 \to d}^{\Delta}
\end{equation}
where
\begin{eqnarray}
\mathcal{M}_{{^1S_{0}} \to d} = 
\left[
G^{IA}_{{^1S_{0}} \to d} +
G^{sea}_{{^1S_{0}} \to d} +
G^{pif}_{{^1S_{0}} \to d} +
G^{\Delta}_{^1S_0 \to d}
\right]
\sqrt{\frac{8 \pi}{3}}
\,
\sum_{i} (1 M_J 1 i |  1 \lambda) Y_{1i}(\hat{\bf q}) 
\end{eqnarray}
To compute the unpolarised cross section from eq.~(\ref{Therm_cross_sec}) one
 has to sum over the final and average over the initial spin directions.
 Carrying out these summations one gets the expression for the differential
 cross section for thermal $np$ capture to the $^3S_{1}-{^3D_{1}}$ deuteron
 state:
\begin{equation}
\label{Therm_cross_sec2}
\frac{d \, \sigma}{d \, \Omega_{\gamma}} = 
\frac{\omega_{\gamma}}{\left| {\bf v}_n \right|}
\frac{e^2}{4(2\pi)^2} 
\left[
G^{IA}_{^1S_{0} \to d} +
G^{sea}_{^1S_{0} \to d} +
G^{pif}_{^1S_{0} \to d} +
G^{\Delta}_{^1S_0 \to d}
\right]^2  
\end{equation}
After angular integration
 and introducing the relative $np$ momentum in the c.m. system of the incident
 channel:
\begin{equation}
p = \frac{1}{2} M_N \left| {\bf v}_n \right| = 
                    \left( \frac{1}{2} M_N E_n \right)^{1/2},
\end{equation}
one arrives at:
\begin{eqnarray}
\sigma (np \to d \gamma) = 
\alpha \frac{\omega_{\gamma} M_N }{2 p}
\left[
G^{IA}_{^1S_{0} \to d} +
G^{sea}_{^1S_{0} \to d} +
G^{pif}_{^1S_{0} \to d} +
G^{\Delta}_{^1S_0 \to d}
\right]^2 \hspace{0.5cm}
\end{eqnarray}
where $\alpha=e^2/4\pi$ is the fine structure constant.

The relative change of the IA result due to the exchange currents 
can be expressed in terms of the quantity
\begin{equation}
\delta(\mathrm{MEC}) = \frac{G^{sea}_{^1S_{0} \to d} +
G^{pif}_{^1S_{0} \to d}}{G^{IA}_{^1S_{0} \to d}} \equiv 
\delta^{sea} + \delta^{pif}
\end{equation}
The empirical value for $\delta$ is normally deduced from the relations:
\begin{equation}
\label{Empir_delta}
\sigma^{exp} (np \to d \gamma) = (1 + \delta)^2 \cdot
\sigma^{IA} (np \to d \gamma)
\end{equation}
where
\begin{equation}
\sigma^{IA} (np \to d \gamma) = 
\alpha \frac{\omega_{\gamma} M_N }{2 p}
\left[
G^{IA}_{^1S_{0} \to d}\right]^2 
\end{equation}
and $G_{^1S_0 \to d}^{IA}$ is determined by eq.~(\ref{IA_res}) with $q =
 \omega_{\gamma}$.  We have now at our disposal all formulae for the analysis
 of both the IA- and the MEC-contributions in the DBS model. All the above
 formulas for transition matrix elements are derived for the general case and
 can be applied to related processes as well.  In the long-wave limit, the
 equations for the MEC will lead to the forms familiar from
 literature~\cite{Mathiot89}.

\section{Results and discussion}
\label{sec:result}
The most important uncertainty in IA comes from the asymptotic behaviour of
 the wave function, because the momentum transfer carried by the photon is
 almost zero.  Therefore, it is not surprising that 
the analysis of the theoretical cross sections within
 IA~\cite{Mathiot89} has demonstrated a very high stability of the IA-results
 for different $NN$-potential models (see Table~\ref{Tot_cross}).

\begin{table}[h]
\begin{center}
\caption{\label{Tot_cross}
The values of the one-body contribution for the different
potentials to the thermal neutron capture reaction as compared 
with experimental data.}
\begin{tabular}{lc|ll}
\hline
\hline
&&\\[-1ex]
potential   & Ref.  & IA  [\mbox{mb}]  & $\delta (\%)$ \\[1ex]
\hline
&& \\[-1ex]
TS          &\cite{Tourreil}  &  305   & 4.68      \\
Argonne V14 &\cite{1a}        &  304.1 & 4.83      \\
Paris       &\cite{Paris_pot} &  302   & 5.2       \\
RSC         &\cite{RSC_pot}   &  300   & 5.55      \\[1ex]
\hline
&& \\[-1ex]
DBS model   & present         &  304.3 & 4.79      \\[1ex]
\hline
&& \\[-1ex]
Experiment  &\cite{CWC}       & 334.2$\pm 0.5$ &  \\[1ex]
\hline
\hline
\end{tabular}
\end{center}
\end{table}

In Table~\ref{Tot_cross} the IA results and the normalisation~$\delta$ to the
 experiment are compiled for the potentials of
 de-Tourreil-Sprung~\cite{Tourreil}, Argonne V14~\cite{1a},
 Paris~\cite{Paris_pot} and Reid Soft-Core~\cite{RSC_pot}, respectively. The
 small variations among different $NN$-model predictions are due to the
 differences in $D$-wave admixture (which changes the value of parameter $A_S$
 of asymptotic behaviour of the deuteron wave function) and due to the
 different short-range repulsion. The potentials are ordered for increasing
 repulsion from the TS to the RSC potential. Thus, the 10\% discrepancy between
 the experiment and the one-body contribution is significant, and cannot be
 accounted for by any change in the nuclear wave function compatible with the
 fit to the accurate $NN$-data.

It is evident from Table~\ref{Tot_cross} that the IA-results for all realistic
 $NN$-potential models and the DBS-model as well agree to each other within
 some narrow corridor.  It should be expected because the long-range behaviour
 of $NN$ wave functions for all the models is quite similar.  However, the
 close similarity of the IA results for all $NN$-models considered here is
 still informative because the DBS wave functions in both initial and final
 channels have a node, and thus the short-distance behaviour is sharply
 distinct from those for the traditional $NN$-potential models.  Thus the
 30~$mb$ residual of the cross section should come from MEC-contributions for
 all $NN$-models, no matter whether they are based on meson-exchange mechanism,
 hybrid six-quark model or on the dressed-bag intermediate state
 production. This is our first conclusion.

The $\delta$-value of eq.~(\ref{Empir_delta}) necessary to match the
 DBS-calculation to the experimental cross section of the $np \to d \gamma$
 reaction amounts to
\begin{equation}
\delta = 4.79 \%
\end{equation}
Our results for the MEC-contributions within the DBS-model which include the
 seagull, meson-in-flight and isobar current terms are presented in
 Table~\ref{MEC_results}. For clarity of the conclusions we show in this Table
 the results for both soft ($\Lambda_{\pi} = 697$~MeV) and hard
 ($\Lambda_{\pi} = 1200$~MeV) $\pi NN$ and $\pi N \Delta$ form factors. It is
 evident that going to hard cut-off parameters increases all the
 MEC-contributions quite visibly, as was expected before. The MEC and IC
 contributions in our calculation with the soft $\Lambda_{\pi NN}$ give in sum
\begin{equation}
\delta (\mbox{MEC + IC}) \simeq 2.8 \%
\end{equation}
and hence there remains some 2\% of the total cross section still unexplained
 with the present $NN$-model. Therefore it is appropriate here to discuss
 which further contributions should appear in our approach.

\begin{table*}[h]
\begin{center}
\caption{ \label{MEC_results}
Relative contributions of seagull, mesonic and model dependent resonance terms
 for pion exchange currents in the DBS model for two cut-off
 factors~$\Lambda_{\pi}$. The values are given for $S\to S$ and $S\to D$
 transitions. 
 The $\delta^{\pi}$ present the values for point-like nucleons.
} \vspace{0.5cm}
\begin{tabular}{l|cc|cc|cc}
\hline
\hline
&&&&&&\\[-2ex]
 Term  &  $\delta^{\pi}_{SS} (\%)$ & $\delta^{\pi}_{SD} (\%)$ &  
$\delta^{\pi + FF \pi}_{SS} (\%)$  &  $\delta^{\pi + FF \pi}_{SD} (\%)$ 
&$\delta^{\pi + FF \pi}_{SS} (\%)$  & 
$\delta^{\pi + FF \pi}_{SD} (\%)$\\[1ex]
 &&&\multicolumn{2}{c|}{$\Lambda_{\pi} = 697$~MeV}&
          \multicolumn{2}{c}{$\Lambda_{\pi} = 1200$~MeV}\\[1ex]
\hline
&&&&&& \\[-2ex]
Seagull & ~2.707 &0.459&~2.120 &0.393&~2.422 & 0.447\\
Mesonic  &-1.243 &0.459&-1.065 &0.393&-1.117& 0.447\\[2ex]
Total MEC&~1.464 &0.918&~1.005 &0.786&~1.305 & 0.954\\[1ex]
\hline
&&&&&& \\[-2ex]
Isobar current & -  & 1.151   &  -   & 0.974 & -  & 1.062\\[1ex]
\hline
&&&&&& \\[-2ex]
Total MEC+IC & ~1.464  &  2.069  & ~1.005 & 1.760 & ~1.305 & 2.016\\[1ex]
\hline
\hline
\end{tabular}
\end{center}
\end{table*}

\begin{figure}[t]
\begin{center}
\includegraphics[clip=true,width=0.35\textwidth]{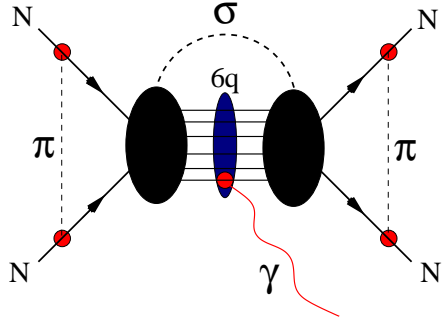}
\caption{\label{6q}
The 6-quark contributions to the DBS e.-m. current induced by
transitions between dressed-bag states.}
\end{center}
\end{figure}

The main difference between the DBS-model predictions and the conventional
 $NN$-potential models comes from the mixed two-component quark-meson
 structure of DBS-approach. The most evident specific contributions are
\begin{enumerate}
\item[i)] The first non-vanishing contribution, which is solely due to the
   two-component structure, are transitions between dressed-bag states
   themselves (these currents can be identified with the operator $\hat j
   ^{6q}_{\mu}$ in eq.~(\ref{DBS_current})).  These can be illustrated for
   $^1S_0 \to (^3S_1 - ^3D_1)$ transitions by the following graphs in
   Fig.~\ref{6q}.
This process corresponds to the $M1$ single-quark transition in the bag. We
 would like to stress the point that the spin-flip of the quark in the bag is
 a non-exotic process because in the conventional quark-diquark model for the
 nucleon, the usual nucleon spin-flip (which is responsible for the
 conventional $(\vec \sigma \cdot \vec \nabla)$- coupling in $\pi N$-
 amplitude) originates also from the unpaired quark spin-flip transition.
\item[ii)]
The dressed six-quark bag in our model has mainly an uncharged $\sigma$-meson
    cloud with the effective mass of the meson being $m_{\sigma} \simeq
    400$~MeV.  The charged six-quark core and the chargeless $\sigma$-cloud are
    oscillating in the intermediate DBS-state around the total
    center-of-mass. Thus these oscillations of the charged six-quark core will
    induce some e.m. radiation similar to radiation emitted by a charged
    oscillator in conventional quantum theory. Therefore this radiation from
    the DBS-component in the $NN$ system may have monopole and also higher
    multipole character.  A $M1$ transition should be especially
    appropriate to describe  polarisation degrees of freedom in the
    $\stackrel{\rightarrow}{n} p \to d \gamma$ reaction. The point here is,
    that the scattering and bound state wave functions in the $NN$~channel in
    our model are non-orthogonal to each other due to the non-hermicity of the
    underlying $NN$-``potential'' $\mathcal{V}_{NqN}(E)$. However, the total
    two- (or many-) component functions $\Psi(NN)$ should still be orthogonal
    to each other and thus monopole $E0$ transitions are quite possible to
    occur {\it separately} in the $NN$- or in the DBS-channel while the
    monopole transition between the initial and final state vectors
    should vanish. The above monopole transition would include the transition
    $^3S_1 \to ^3S_1$ in the deuteron. However, the isoscalar $M1$-transition
    from the {initial} $^3S_1$ channel to the final $^3S_1$ channel inside the
    bag, which is related to the spin rotation of the bag as the whole from
    $M_S = +1$ (or -1) to $M'_S=0$ (or vice versa), is still feasible.
\item[iii)]
    The dressing of the bag includes in our model not only neutral $\sigma$,
    but also charged $\pi$- and $\rho$-meson clouds. It would be quite
    possible for incoming $\gamma$-quanta to interact first with these charged
    clouds of the bag. These specific new meson-exchange currents will
    contribute to the $np \to d \gamma$ reaction and also to many other
    e.m.
    processes~\cite{Kaskulov:an,Obukhovsky:ae,Obukhovsky:2002xn}. 
    We postpone the discussion of these new effects to our
    further publications.
\end{enumerate}

In summary, we presented in this paper the first e.-m. calculations for $np
 \to d \gamma$ cross sections on the basis of the new $NN$-interaction
 model. The model employs the soft meson-nucleon cut-off parameter values so
 that we still observe some underestimation of the experimental cross
 section. However, there are a few specific new contributions in the model which
 are fully absent in conventional $NN$-potential models. Our further
 studies point in this direction.

~\\ 
\noindent{\bf Acknowledgement}

The authors are thankful to many of our colleagues for fruitful discussions
 and help. We thank especially Drs. I.T.~Obukhovsky and V.N.~Pomerantsev for
 their numerous advices and informal discussion. One of the author V.I.K. is
 grateful to the Russian Foundation for Basic Research 
 (Grants No 02-02-16612, 01-02-04015),
 and the Deutsche Forschungsgemeinschaft (Grant No Fa-67/20-1) for partial
 financial support of his work; M.M.K. is grateful to the Deutsche
 Forschungsgemeinschaft for financial support of this work (Gr~1084/3-3).

\appendix\section{One- and Two-Pion Exchange Potentials in the DBS-model}
\label{sec:app}

If isospin-breaking terms are ignored and staying within the lowest order of
 perturbation theory, the OPE $NN$- potential in momentum space takes the
 form:
\begin{equation}
V_{ij}^{OPE}({\vec q}) = - \frac{f_{\pi NN}^2}{m^2_{\pi}} 
\frac{({\vec \sigma}_i \cdot {\vec q}) 
({\vec \sigma}_j \cdot {\vec q})}{{\vec q}^2 + m^2_{\pi}}
({\vec \tau}_i \cdot {\vec \tau}_j),
\end{equation} 
where  $m_{\pi}$ is the mass of the exchanged pion and $f_{\pi NN}^2$ the 
coupling constant.

When the potential is Fourier
 transformed to configuration space, it is usually regularised to remove
 the singularities at the origin. This can be achieved by introducing  
 a form factor $F({\vec q}^2)$ which has a dipole form:
\begin{equation}
F({\vec q}^2) = \left( \frac{\Lambda_{\pi NN}^2 - m_{\pi}^2}
{\Lambda_{\pi NN}^2 + {\vec q}^2} \right)^2
\end{equation}
where $\Lambda_{\pi NN}$ is the cut-off parameter. 

A typical Fourier transform reads as
\begin{equation}
\label{OPEP_Fourier}
\int \frac{d^3 {\vec q}}{(2 \pi)^3} \frac{e^{i{\vec q}{\vec r}}}{{\vec q}^2 + m^2_{\pi}} 
({\vec q}^2)^n F({\vec q}) \equiv \frac{m_{\pi}}{4 \pi}
  (- {\vec \nabla}^2)^n f_C(r)\ .
\end{equation}
So, in configuration space the OPE potential takes the form:
\begin{equation}
V_{ij}^{OPE}({\vec r}) = \frac{f_{\pi NN}^2}{4 \pi} \frac{m_{\pi}}{3} \left[ 
f_C(r)
{\vec \sigma}_i \cdot {\vec \sigma}_j +  
f_T(r)
S_{ij} \right] {\vec \tau}_i 
\cdot {\vec \tau}_j\ ,
\end{equation}
where  the tensor interaction operator is
\begin{equation}
S_{ij} = 3 (\vec{\sigma} _i \cdot \hat{\vec r}) (\vec{\sigma} _j
  \cdot \hat{\vec r})
- (\vec{\sigma} _i \cdot \vec{\sigma} _j)
\end{equation}
and the central interaction term is
\begin{equation}
f_C(r) = \displaystyle 
\left[ \frac{e^{- x}}{x} - 
\frac{e^{- \alpha x}}{x} \right] -
e^{- \alpha x}
\left( \frac{\alpha^2 - 1}{2} 
\alpha \right)\ ,  
\end{equation}
with
\begin{equation}
x = m_{\pi} r\ \ \ \mathrm{and}\ \ \  \alpha = \frac{\Lambda_{\pi NN}}{m_{\pi}}
\ .
\end{equation}
Using the definition (\ref{OPEP_Fourier}), the Fourier transform for the
tensor potential can be  expressed simply in  
terms of derivatives of the central potential, i.e.,
\begin{equation}
f_T(r) = 
\frac{1}{ m_{\pi}^2} r \frac{d}{d \, r} \left( \frac{1}{r}
 \frac{d}{d\, r} \right) f^0_C(r)\ ,
\end{equation}
with
\begin{equation}
f^0_C(r) = \displaystyle 
\left[ \frac{e^{- x}}{x} - 
\frac{e^{- \alpha x}}{x} \right] -
e^{- \alpha x}
\left( \frac{\alpha^2 - 1}{2 \alpha} 
\right)\ ,  
\end{equation}
and the tensor interaction term can be written in terms of 
the variables $x$ and $\alpha$ as
\begin{eqnarray}
f_T(r) =  \frac{e^{-x}}{x} \left( 1 + \frac{3}{x} + \frac{3}{x^2} \right) 
-
\alpha^3 \frac{e^{- \alpha x}}{\alpha x} 
\left[ 1 + \frac{3}{\alpha x} + \frac{3}{(\alpha x)^2} \right] - 
\frac{\alpha}{2} \left( \alpha^2 - 1 \right) \left[1 + \frac{1}{\alpha x}
 \right] e^{- \alpha x}\ . \nonumber \\
\end{eqnarray}
We use  the averaged pion mass in the OPE potential:
$$m_{\pi} = (m_{\pi^0} + 2 m_{\pi^{\pm}})/3 = 138.0363 \,\, \mathrm{MeV},$$
and the respective ``weighted'' value for the pion-nucleon coupling constant
$$f^{2}_{\pi NN}/(4 \pi) = 0.075$$
The value of the regularisation parameter $\alpha$ amounts to
$$\alpha = 5.054.$$

The inclusion of the peripheral two-pion exchange contribution~$V_{TPE}$ in
 our case can be imitated by the simplified form~\cite{4a}:
\begin{equation}
V_{TPE} = V_{TPE}^0 (\beta r^2)^2 e^{-\beta r^2}
\end{equation}
and gives only a quite small contribution (about 2-3~MeV) in the intermediate
 region $r \sim 1.5 \div 2$~fm which is well beyond the DBS radius. However,
 this small attractive contribution is quite important for the precise
 description of effective-range parameters and asymptotic properties of the
 deuteron.

\end{document}